\documentclass[acmtog,nonacm]{acmart}

\AtBeginDocument{%
  \providecommand\BibTeX{{%
    \normalfont B\kern-0.5em{\scshape i\kern-0.25em b}\kern-0.8em\TeX}}}

\AtBeginDocument{%
  \fancypagestyle{plain}{%
    \fancyhf{} %
    \fancyfoot[L]{ACM SIGENERGY Energy Informatics Review}%
    \fancyfoot[R]{Volume 1 Issue 1, November 2021}}
  \providecommand\BibTeX{{%
    \normalfont B\kern-0.5em{\scshape i\kern-0.25em b}\kern-0.8em\TeX}}}
    
\let\oldmaketitle\maketitle
\renewcommand{\maketitle}{%
  \oldmaketitle%
  \thispagestyle{plain}%
  \pagestyle{plain}}
\begin{document}

\title{A Holistic Review on Advanced Bi-directional EV Charging Control Algorithms}

\author{Xiaoying Tang}
\affiliation{%
  \institution{School of Science and Engineering, The Chinese University of Hong Kong (Shenzhen), and the Shenzhen Institute of Artificial Intelligence and Robotics for Society}
  \city{Shenzhen}
  \country{China}}
    \thanks{This work is
supported in part by the funding from Shenzhen Institute of Artificial Intelligence and Robotics for Society, the National Key R\&D Program of China with grant No. 2018YFB1800800, and the National Natural Science Foundation of China (NSFC) under Grant No. 62001412.} 
\email{tangxiaoying@cuhk.edu.cn}

\author{Chenxi Sun}
\affiliation{%
  \institution{School of Science and Engineering, The Chinese University of Hong Kong (Shenzhen), and the Shenzhen Institute of Artificial Intelligence and Robotics for Society}
  \city{Shenzhen}
  \country{China}}
\email{sunchenxi@cuhk.edu.cn}

\author{Suzhi Bi}
\affiliation{%
  \institution{Shenzhen University}
  \city{Shenzhen}
  \country{China}}
\email{bsz@szu.edu.cn}

\author{Shuoyao Wang}
\affiliation{%
  \institution{Shenzhen University}
  \city{Shenzhen}
  \country{China}}
\email{sywang@szu.edu.cn} 

\author{Angela Yingjun Zhang}
\affiliation{%
  \institution{The Chinese University of Hong Kong}
  \city{Hong Kong}
  \country{China}}
\email{yjzhang@ie.cuhk.edu.hk}  
  

\begin{abstract}
The rapid growth of electric vehicles (EVs) has promised a next-generation transportation system with reduced carbon emission.
The fast development of EVs and charging facilities is driving the evolution of Internet of Vehicles (IoV) to Internet of Electric Vehicles (IoEV).
IoEV benefits from both smart grid 
and Internet of Things (IoT) technologies  which provide advanced bi-directional charging services and real-time data processing capability, respectively. 
The major design challenges of the IoEV charging control lie in   the randomness of charging events and the mobility of EVs.
In this article, we present a holistic review on advanced bi-directional EV charging control algorithms.  For Grid-to-Vehicle (G2V), we introduce the charging control problem in two scenarios: 1) Operation of a single charging station and 2) Operation of multiple charging stations in coupled transportation and power networks. 
For Vehicle-to-Grid (V2G), we discuss how EVs can perform energy trading in the electricity market and provide ancillary services to the power grid.
Besides, a case study is provided to illustrate the economic benefit of the joint optimization of routing and charging scheduling of multiple EVs in the IoEV. Last but not the least,
we will highlight some open problems and future research directions of charging scheduling problems for IoEVs.
\end{abstract}



\maketitle

\section{Introduction}
As an environmental friendly substitute for traditional fuel-powered vehicles, electric vehicles (EVs) lie at the heart of future sustainable and smart transportation systems. 
The rapid development of EVs and charging facilities is driving the evolution of Internet of Vehicles (IoV) to Internet of Electric Vehicles (IoEV).
However,  uncontrolled EV charging can result in expensive power generation cost, transmission congestion, and even cause security issues to the smart grid \cite{muratori2018impact}.

The recent development of smart grid technology provides a new set of tools that enable more secure and efficient EV charging.
For instance, advanced charging facilities enable both Grid-to-Vehicle (G2V) and Vehicle-to-Grid (V2G) power flows such that EVs can act as not only electricity load consumers, but also energy providers that compensate for the power deficiency in peak load hours.
Besides, EVs can act as mobile energy storage to transport excessive energy generated by remote renewable sources to the main grid \cite{lam2018opportunistic}.
Furthermore, the emerging Internet of Things (IoT) technology provides a platform to control various loads and manage the charging facilities \cite{sheng2018toward}.
By supporting rapid and secure data collection, distribution, and information exchange, IoT technology enables many advanced data processing and performance optimization technologies in  power grids and city transportation systems that would significantly enhance the EV charging efficiency.
\begin{figure*}
\centering
\includegraphics[width=0.6\textwidth]{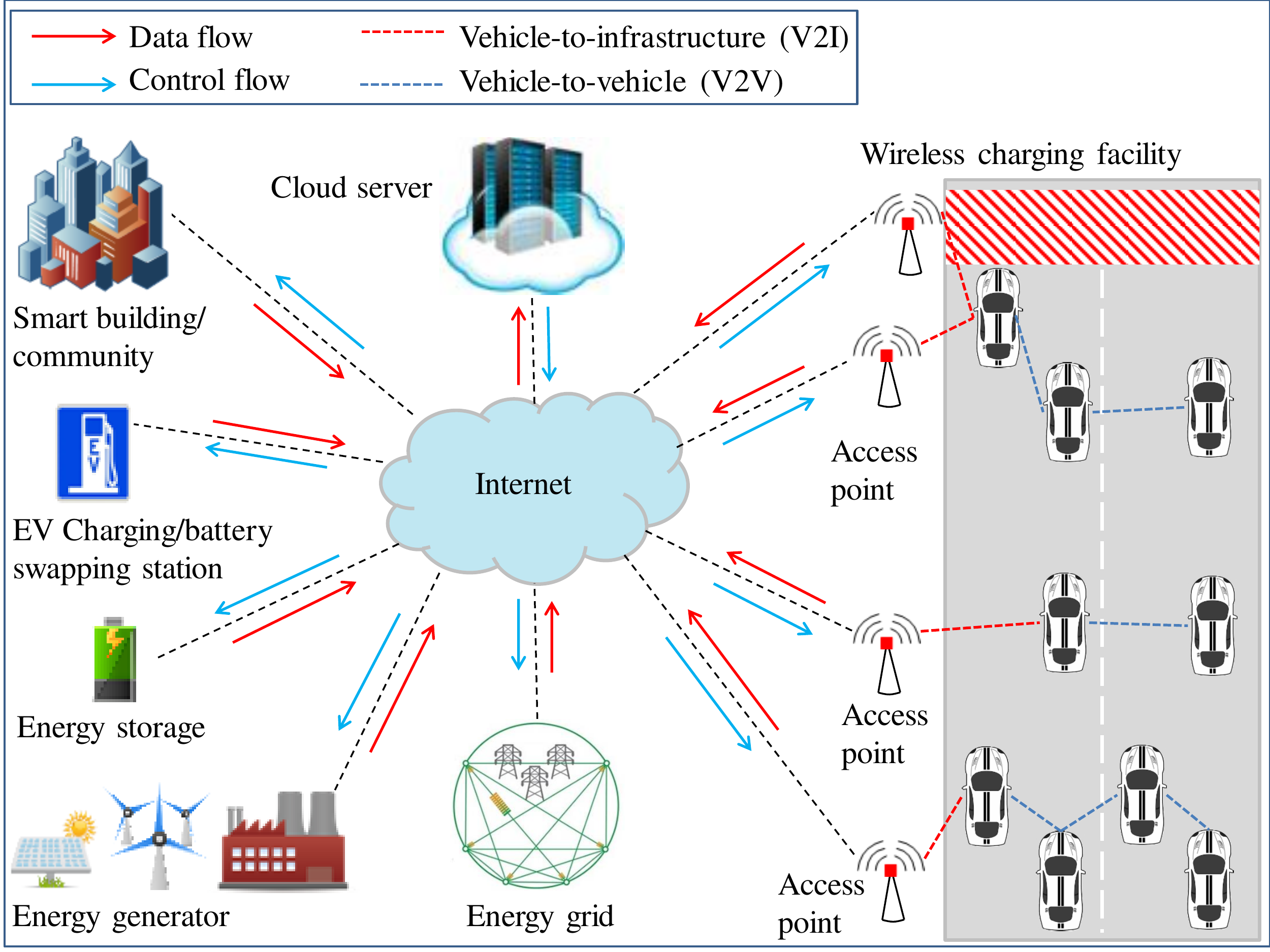}
\caption{Illustration of IoEVs and related energy systems. IoT  interconnects  a  massive  number of EVs, charging facilities and other critical components that affect  the  performance  of  the  IoEVs.}
\label{fig:IoT:EVN}
\end{figure*}

\begin{figure*}
\centering
\includegraphics[width=0.6\textwidth]{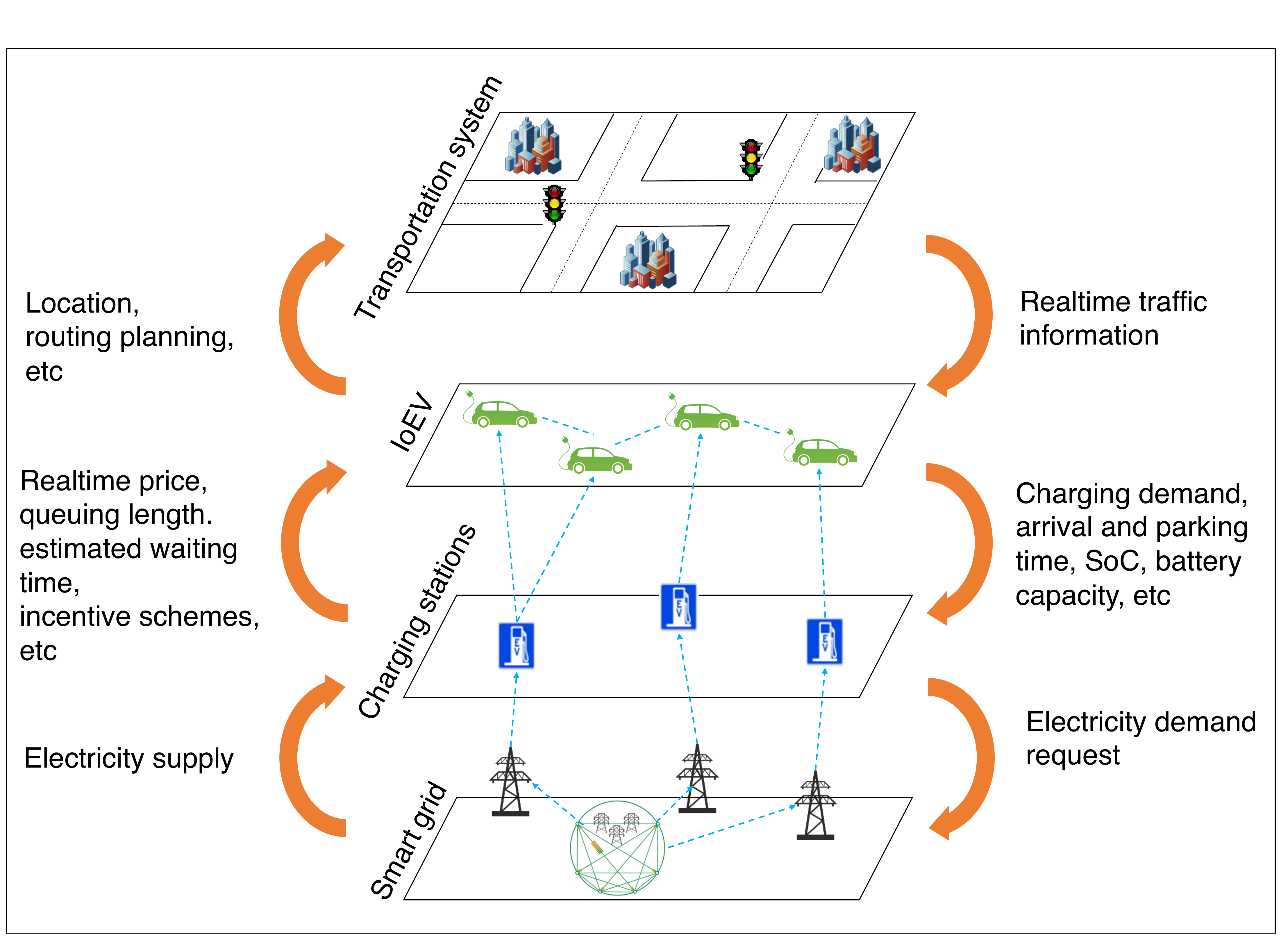}
\caption{An overview of multi-layer system and interaction process of IoEV, smart grid, charging stations and transportation system.  IoEV collect not only information from  charging  stations  at  different  locations,  but  also  real-time traffic from the intelligent transportation system. In the meantime, IoEV can send charging information to charging station of interest and update the routing planning to the transportation system.   }
\label{fig:IoEV:multi-layer}
\end{figure*}

As shown in Fig.~\ref{fig:IoT:EVN}, IoT interconnects a massive number of EVs, charging facilities and other critical components that affect the performance of the IoEVs. 
For instance, different types of energy generators lead to different location-dependent electricity prices; the power consumptions of commercial and residential buildings cause time-varying electricity prices; energy storage can absorb the excess energy generated from renewables to meet energy deficiency of the EVs during peak hours.
In particular, 4G/5G, IEEE 802.11p and other wired/wireless communication technologies enable real-time data collection and information exchange among different components, such as pricing and congestion conditions.
For instance, EVs can operate in a vehicle-to-vehicle (V2V) mode, where IEEE 802.11p is used to exchange road conditions with neighboring EVs, or in a vehicle-to-infrastructure (V2I) mode, where LTE/4G is used to receive  electricity charging prices updated and broadcasted by the system operator.
Due to the resource constraints of IoEVs, data processing is often delegated to cloud servers with strong computational power.
As such, mobile edge/cloud computing enhances the functionality of IoEVs in terms of data storing, processing, dissemination and fast computation.

The routing and charging behaviour of EVs leads to real-time interactions between IoEV, smart grid, charging stations, and transportation system. 
As shown in Fig.~\ref{fig:IoEV:multi-layer}, IoEV can collect not only the information about price and waiting time from charging stations at different locations, but also the real-time traffic information from the intelligent transportation system. Based on the received information, each EV makes sequential decisions for charging/discharging and routing planning, and informs the charging station of interest about its status, such as charging demand, estimated arrival and parking time, battery capacity, etc. The EV may also update its information, such as location and route planning, to the transportation system.
In the meantime, based on the received information from IoEV, charging stations can update the pricing or incentive schemes to maximize the system utilities, and update the electricity requests to the smart grid.
In general, the challenges of bi-directional IoEVs charging control lie in two aspects. 
The first challenge is due to system randomness, including the random
charging profiles of EVs (arrival, departure, charging demand,  state of charge (SoC), etc.), random future load demand in the main grid, random renewable energy generations and random electricity prices.
The other challenge is due to the coupling effect between the
smart grid and transportation network in the sense that an EV can only replenish or discharge its battery at charging stations in its route. 
As such, IoEV charging scheduling, charging station selection, and routing decisions are strongly coupled.

In the remainder of the paper, Section \ref{g2v} presents the problem setting and charging control techniques for a single charging station and for multiple charging stations in coupled transportation and power networks.  
Section \ref{v2g} describes how EVs can perform energy trading in the electricity market and provide ancillary services to the power grid.
Section \ref{sec:case study} provides a case study to illustrate the economic benefit of the joint optimization of routing and charging scheduling of multiple EVs in the IoEV.
Section \ref{dis} highlights some open problems and future research directions of efficient charging control for IoEVs.
Lastly, we draw conclusions in Section \ref{con}.

\section{Grid-to-Vehicle (G2V)} \label{g2v}
\subsection{Operation of a single charging station}

\begin{figure}
\centering
\frame{\includegraphics[width=0.9\linewidth]{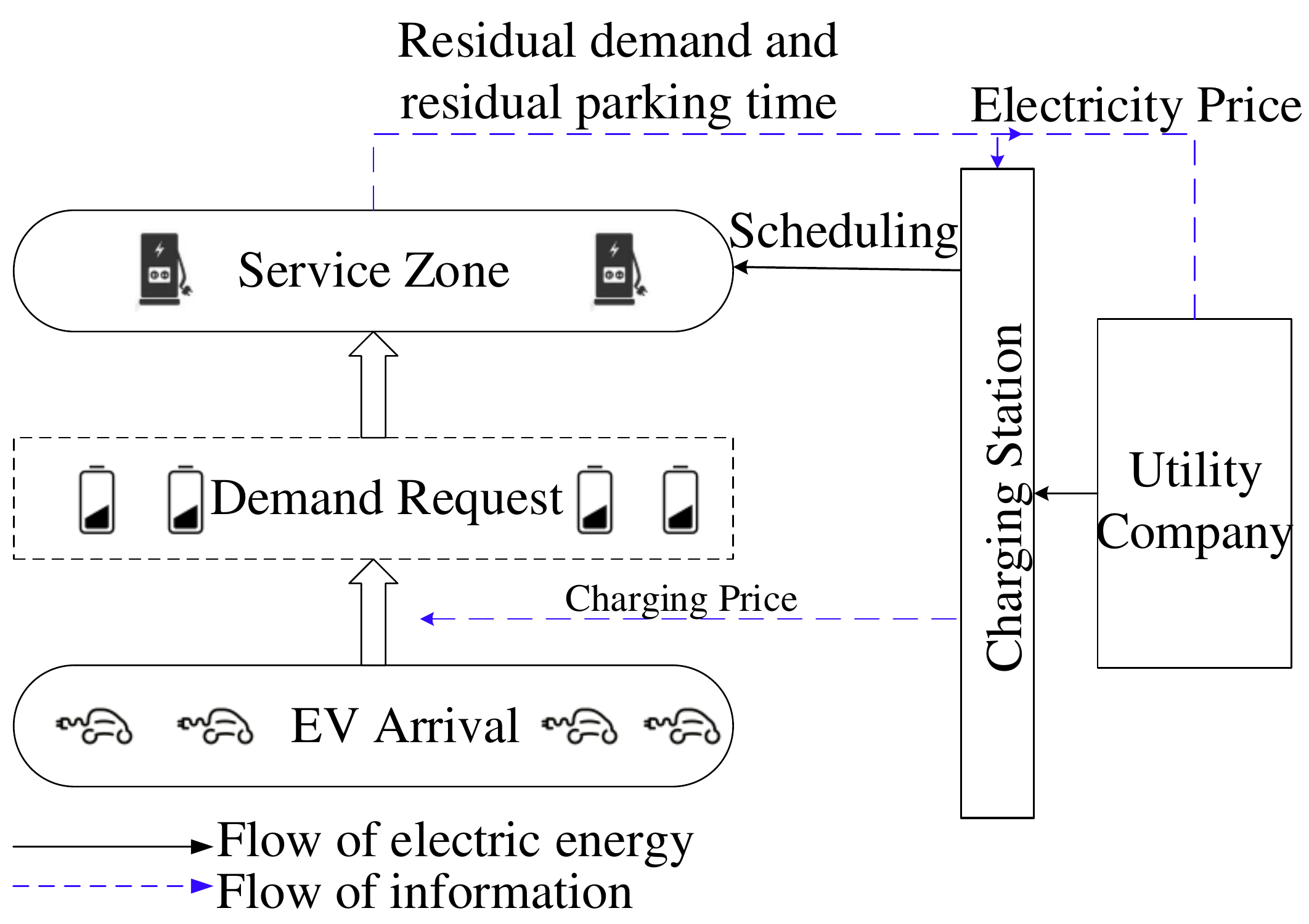}}
\caption{The interaction process of charging station, EVs and utility company.}
\label{fig:StationModel}
\end{figure}

A general interaction process of charging station, EVs and utility company is shown in Fig.~\ref{fig:StationModel}. EVs arrive at the charging station randomly, and each EV expects the charging station to fulfill its charging demand within an expected time period. 
At each time slot, the charging station receives the current electricity price from the utility company and broadcasts a charging price to all arriving EVs.
When an EV arrives,  it attempts to minimize its charging cost by determining its
charging demand according to the charging price, and then sends the demand request to the charging station. 
Based on the requested demand, the charging station decides
whether to admit the EV in order to avoid excessive delay of admitted EVs.  
Once admitted, the EV enters the service zone where there are several parking lots and charging ports. 
Then, the charging station schedules the charging power/rate for each EV plugged in the charging ports and collects charging fees from the EV users accordingly.
In some practical scenarios, the number of charging ports may be smaller than that of the admitted EVs, so that the EVs may have to wait for charging after being admitted.
In this case, the waiting time of the admitted EVs is sent back to the controller of the charging station.
The charging station has the following control methods to optimize the charging performance. 


\emph{Pricing Scheme}:
Pricing is a type of demand-response mechanisms where EVs adjust their charging demands according to the charging price announced by the charging stations.
Therein, the charging station can design the pricing scheme to control the charging demands of EVs to maximize its overall profit and system efficiency.

\emph{Admission Control}:
An admission control policy refers to the policy where the charging station selectively admits EV's charging requests. Admission control is very useful to  reduce the charging waiting time.
Here, the charging waiting time is defined as the time between the
arrival time of an EV and the time that the EV starts to receive service.
The charging capacity of a charging station is limited by two factors: 1)  the total charging power of a charging station bounded due to physical and security constraints of the distribution network; 2) the number of EVs that a charging station can accommodate limited by the hardware and space constraints. 
A long charging waiting time degrades the users’ experience, and further negatively impacts the charging station's long-term profit.
In practice,  a waiting time penalty should be considered in the system model. 
A naive admission control method is the queue-length based admission (QBA) policy, where a newly arrived EV is admitted only if the number of EVs waiting to be served at the station is below a specific threshold. 
However, it may perform  poorly due to the negligence of the user demand differences \cite{wang2018electrical}.
Therefore, a good admission control design should be  based on the actual charging demands brought by the EVs.

\emph{Scheduling}:
Charging scheduling refers to the sequential decisions made by charging stations on how much power to charge each admitted EV at each time. 
The decisions are generally made based on the past and current information of EVs that have already arrived. 
There exists some greedy charging scheduling methods that maximize the current revenue without considering the unknown future charging demands.
These approaches may suffer a high penalty in the future, e.g., paying higher electricity price or penalty because of low service quality \cite{tang2014online,tang2016online}.
In contrast, a good online charging scheduling decision should take into account the random future events, which include the demand, time of arrival and departure of EVs, elastic and inelastic load demand in the power system, renewable generations, realtime electricity prices and regulation service prices, etc.

In recent years, many charging scheduling algorithms have been developed under a variety of settings.
\cite{lee2020adaptive} presented a flexible adaptive scheduling algorithm based on convex optimization and model predictive control and allows for significant over-subscription
of electrical infrastructure. 
\cite{long2021efficient} studied the real-time operation of a public charging station providing charging service to large-scale Plug-in Electric Vehicles (PEVs).
\cite{zeballos2019proportional}  introduced a new policy called least laxity ratio to achieve a
suitable notion of proportional fairness.
\cite{vsepetanc2021cluster} proposed an operating model that can be used both for the day-ahead scheduling and for the intraday model-predictive-control-based adjustments, assuming that both the charging stations and the EV fleets belong to the same company.
\cite{madahi2020co} proposed a new day-ahead co-optimization algorithm to reduce the detrimental effects of PEVs on the power system.
\cite{liu2020optimal} proposed an optimal charging scheduling method that minimizes the operation cost by responding to the time-of-use (TOU) electricity price.

Some existing studies investigated the joint optimization of pricing and the charging scheduling schemes that benefit both EV users and the charging stations. In this case, both the charging rate and charging price are control variables of the charging stations. For example,  \cite{wang2021rein} formulated the pricing and scheduling problem into an Markov decision process and proposed a reinforcement learning approach that maximizes the profit of a charging station.
There are also some studies that jointly optimize pricing and admission control schemes to maximize the total profit of charging station as well as minimize the waiting time of EV users.
In this case, the control variables include the charging price, the number and total charging demands of admitted EVs, while generally simple charging schedule schemes are adopted, e.g.  first come first serve with constant charging/discharging rates.
The key point of jointly optimizing pricing and admission control schemes is to strike a good balance among the waiting time, admission probability, and charging port utilization.
For example, \cite{wang2018electrical} 
analyzed the EV queueing dynamics and derived the waiting time in closed-form, and accordingly proposed a novel multi-sub-process based admission control scheme in order to jointly optimize the profit of charging stations and the delay of EV users.

Some work \cite{wang2018two,fallah2020charge,zeng2021inducing} considered both admission control and charge scheduling strategy.
\cite{fallah2020charge} formulated a multi-stage stochastic programming model to minimize the expected total energy costs over the finite time horizon.
\cite{wang2018two} proposed a two-stage admission and scheduling mechanism to find the optimal tradeoff between accepting EVs and missing charging deadlines under several energy supply scenarios.
\cite{zeng2021inducing} proposed an innovative station-level optimization framework to operate charging station with optimal pricing policy and charge scheduling.

Besides, in order to tackle the system dynamics and randomness of user behavior, data-driven model is another popular method used in energy management for a single charging station. For instance,
\cite{moghaddam2019coordinated} presented a new coordinated dynamic pricing model to reduce the overlaps between residential and charging station loads by inspiring the temporal PEV load shifting during evening peak load hours.
\cite{da2019coordination} proposed a multi-agent multi-objective reinforcement learning architecture that aims at simultaneously minimizing energy costs and avoiding transformer overloads, while allowing EV recharging.
\cite{li2019constrained} proposed a charging scheduling strategy using a safe deep reinforcement learning approach to minimize the charging cost as well as guarantee the EV can be fully charged. 
\cite{wang2019chance} proposed a two-stage energy management system for power grids with massive integration of EVs and renewable energy resources. 
 In \cite{sadreddini2021design}, a smart reservation system considering the
behavior of EV users, parking slot availability, SoC value of EVs, and the parking lot usage history of EV users was proposed.
This line of research utilized historical data such as load, usage history and smart meter measurements to develop effective models for charging station operation.

\subsection{Operation of charging stations (in coupled transportation and power networks)}


In this section, we introduce the joint optimization of routing and charging scheduling for IoEV operation control in transportation network equipped with multiple heterogeneous charging stations.   

\begin{figure}
    \centering
    \includegraphics[width=0.9\linewidth]{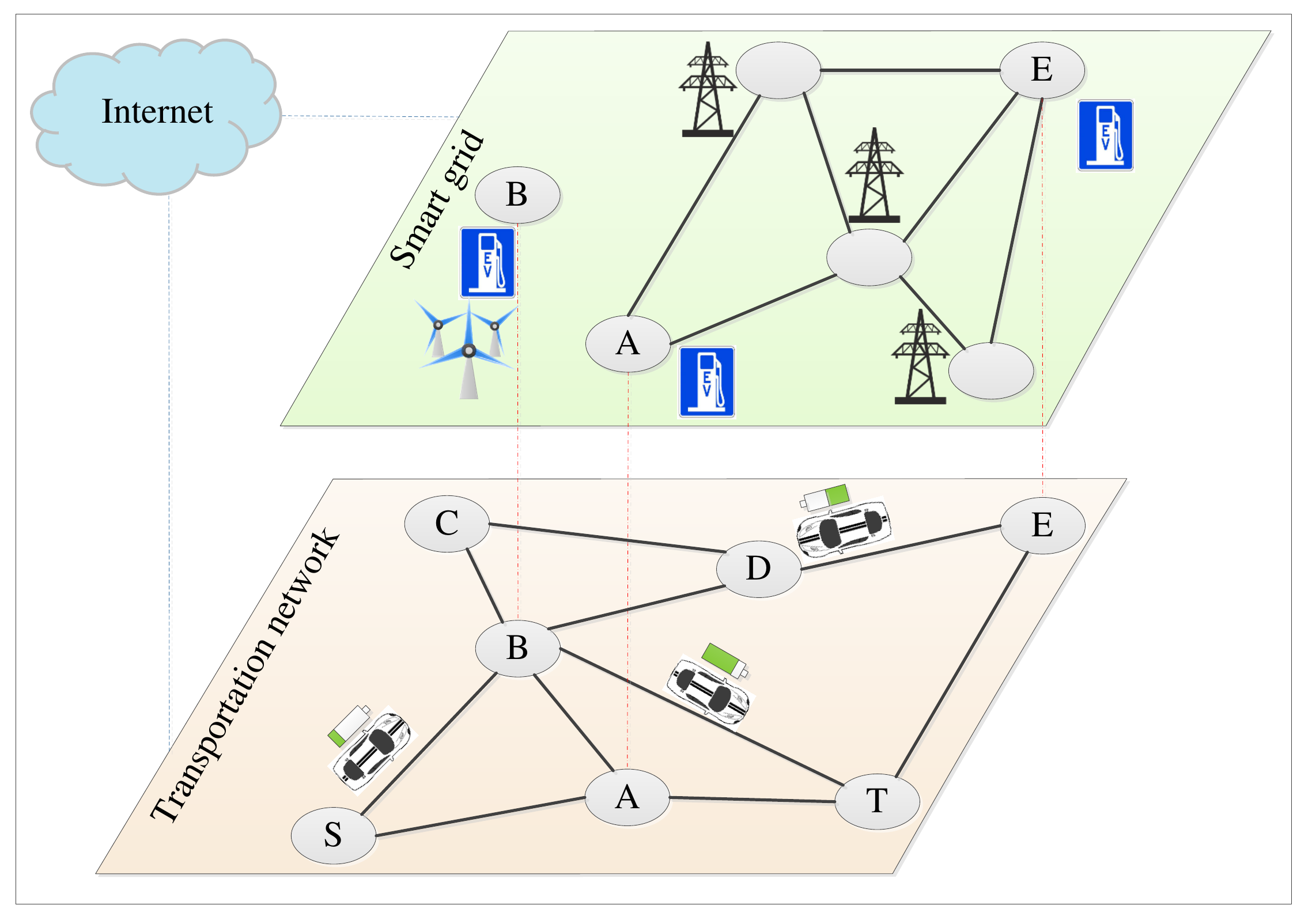}
    \caption{An illustration of the IoEV system coupled with both the smart grid and transportation network}
    \label{fig:routing}
\end{figure}
Optimal routing is a classic problem in conventional transportation networks that aims to minimize travel time, traversed distance and/or energy consumption, etc.
The optimal routing problem is mostly modeled as a shortest path problem in a graph, with some variations of edge weights to consider road congestions, regulations and user preferences.
Commonly used methods include shortest path Dijkstra algorithms, $A^*$ based-search algorithm, Ant Colony optimization, Particle Swarm Optimization, and Tabu Search \cite{adulyasak2016models}.

The conventional routing algorithms cannot be directly adopted to IoEV.
Due to the coupling effect between the smart grid and transportation network, 
the route selection is also coupled with the charging station selection along the selected path, and thus related to the operation of the power system.
Specifically, unlike gasoline price,  electricity price can be significantly different at different charging facilities. For instance, as shown in Fig.~\ref{fig:routing}, the electricity price at node $B$ with renewable energy source is  likely to be much cheaper than that at node $A$, which is powered by the main grid.
Therefore, an EV user may have the incentive to take a detour to charge its battery at station B instead of taking the shortest path.
Besides, the V2G technology allows an EV to sell energy back to the grid for profit.
Under this condition, an EV user may consider the potential profit in route selection by first charging at stations with a lower electricity price (e.g., node $B$) and then discharging its battery at stations with a higher electricity buying price (e.g., node $A$).
Moreover, the electricity price and availability of charging facilities are related to other electricity consumers, such as commercial users and households. For instance, renewable energy is scarce in urban areas with high household consumption but abundant in suburb areas.
As a result, EV routing must be jointly optimized with charging station selection by taking into account all the elements in the system.

In Fig.~\ref{fig:routing}, we use a simple example to show how the electricity prices and V2G technologies influence the routing and charging scheduling of an EV.
Suppose that an EV travels from node $S$ to node $T$. Node $B$ is powered by renewables and node $A$ can buy energy from the EVs.
We assume that the initial battery level of a tagged EV is sufficient to complete both paths $S-A$ and $S-B$.
To maximize the profit of the EV, the optimal routing is $S-B-A-T$ and the optimal charging scheduling is to fully charge the EV at node B and discharged (sell the extra electricity) at node A given that the EV can complete each road segment in the path.
On the other hand, if V2G is not available at node A, the EV will choose the path S-B-T with the minimum energy cost.
In addition, if renewable energy is not available at node B, the EV will select S-A-T with the shortest path and minimum energy consumption.

The problem becomes much more complicated when a large number of EVs plan their routes at the same time.
Uncoordinated planning may lead to overwhelming charging demands at bottleneck charging stations.
Therefore, it is necessary to coordinate the route selection and charging scheduling of EV users to maximize the system performance.
In this case, a major challenge is to design the right incentive scheme, so that the EVs' and system operators' selfish decisions are also the maximizer of the social welfare.
Besides, it is desirable to find scalable algorithms to solve the large-scale routing and charging scheduling problems with affordable communication and computation overheads.

One way to jointly optimize the routing and charging scheduling of a single EV is to model the problem as an extended transportation graph and find a shortest path \cite{alizadeh2014optimized}.
In the case of multi-EV coordination, the decisions of individual EV users are coupled due to the constraint of limited traffic and charging station capacity.
There are two types of control schemes for multi-EV coordination, namely, centralized schemes \cite{Cerna2017optimal} and distributed schemes \cite{cao2018mobile,tang2019distributed}.
Notice that centralized schemes require the EV users to submit their complete information, resulting in serious privacy concerns.
In contrast, distributed algorithms only require little information exchange between the EVs and the system operator, thus significantly reducing the privacy leakage and the complexity of computation and communication compared with centralized schemes.
For instance, \cite{tang2019distributed} proposed a proximal method based distributed algorithm, where the EV users are not required to share their specific route selection with the system operator. 



Some recent studies investigated the joint routing and charging problem from a social coordinator's perspective \cite{sun2020orc,zhang2020power, qian2020enhanced} where the EV owners aimed to find their own optimal charging station and the social coordinator designed pricing strategies such as congestion tolls and locational marginal prices (LMP) to influence or guide EV charging and routing behaviors. 
In \cite{qian2020enhanced}, a bi-level model was proposed to determine the optimal charging service fees for guiding EVs and minimizing the social cost that includes the total driving time, waiting and charging times in transportation network, and total generation cost in power network. 
\cite{sun2020orc} proposed an online recommendation and charging schedule algorithm with on-arrival commitment for sequential EV arrivals that aims to maximize the expected total revenue of a charging station network.
\cite{zhang2020power} adopted an expanded transportation network model to describe transportation constraints and the AC power flow model to describe electrical constraints and proposed a second order cone programming model to minimize the total social cost that includes driving and charging time costs of PEV drivers and power supply costs.

Some studies considered a Charging Network Operator (CNO)'s perspective \cite{elghitani2020efficient,moradipari2019pricing,moghaddass2019smart} where the EV owners cannot directly choose the charging station but are rather assigned to certain stations by a central controller based on the optimization objectives. In this setting, users can specify their desired SoC and their destinations to the CNO and the CNO will assign each EV to an optimal charging station based on the charging request. Specifically, \cite{elghitani2020efficient} developed an EV assignment algorithm based on the Lyapunov optimization method that aims to minimize the average time spent from requesting the service to accessing it. \cite{moghaddass2019smart} formulated an integer multi-objective optimization problem for optimal coordination of a fleet of cooperative EVs considering the objectives of EV owners, charging station owners, and power systems. \cite{moradipari2019pricing} designed pricing and routing policies that ensure users reveal their true needs to the CNO and directly assigned them to a station on their path in order to manage their effects on the grid and ensure fair services. 
This line of work formulated the charging control of EVs in a charging station network as a decision problem of the CNO and focused on developing appropriate methods to find the optimal solution.


\section{Vehicle-to-Grid (V2G)}\label{v2g}

\subsection{Energy trading in the electricity market}
\begin{figure}
    \centering
    \includegraphics[width=0.9\linewidth]{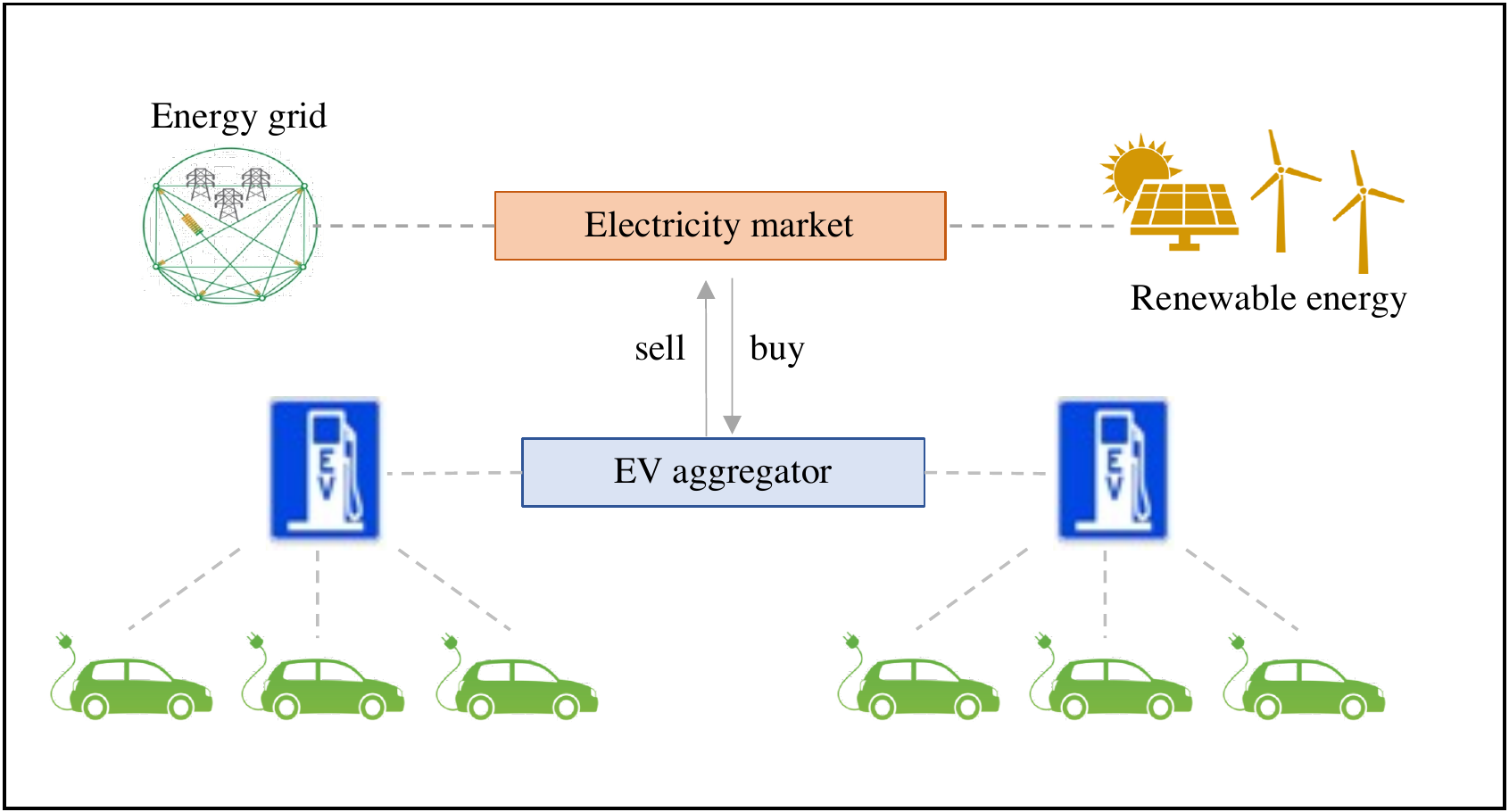}
    \caption{An illustration of EVs participating in the Electricity Market}
    \label{fig:market}
\end{figure}

With the implementation of Vehicle-to-Grid (V2G) technology, EVs can also provide energy to the grid \cite{kempton2005vehicle}. Fig.~\ref{fig:market} gives an illustration of EVs participating in the electricity market. Specifically, due to the limited battery capacity of each EV, EV aggregator (EVA) is normally required to coordinate a collection of EVs in order to participate in the electricity market with bids to purchase or sell electricity. There are two types of electricity/energy market: day-ahead market which let market participants buy or sell electricity one day before the operating day, and real-time market which let market participants buy or sell electricity during the operating day.

Most existing studies \cite{porras2020efficient,han2020optimal,asrari2019market,hajebrahimi2020scenario,liu2021optimal} considered the EVA operation in day-ahead electricity market, as EVA should plan the charging scheduling for each EV beforehand. Specifically, 
\cite{han2020optimal} proposed an optimal operation strategy for an EVA, which performs energy arbitrage in the energy market and provides ancillary services from aggregated EVs, while providing charging services to EVs to maximize the profit in a future energy market.
\cite{porras2020efficient} proposed a hierarchical optimization approach to represent the decision-making of this aggregator in the day-ahead electricity market. 
\cite{asrari2019market} proposed a day-ahead market framework for congestion management in smart distribution networks considering collaboration among EVAs.
In \cite{hajebrahimi2020scenario}, a new distributionally robust optimization (DRO) via scenario wise ambiguity set is proposed to develop a collaborative bidding strategy for intermittent resources such as EVA in the day-ahead energy market.
\cite{liu2021optimal} considered the EVA participation not only in the electricity market, but also reserve market.
These studies investigated the optimal day-ahead operation strategy of EVA and the potential benefit of collaboration among EVAs.

\subsection{Ancillary Service by EVA}

\begin{figure}
    \centering
    \includegraphics[width=0.9\linewidth]{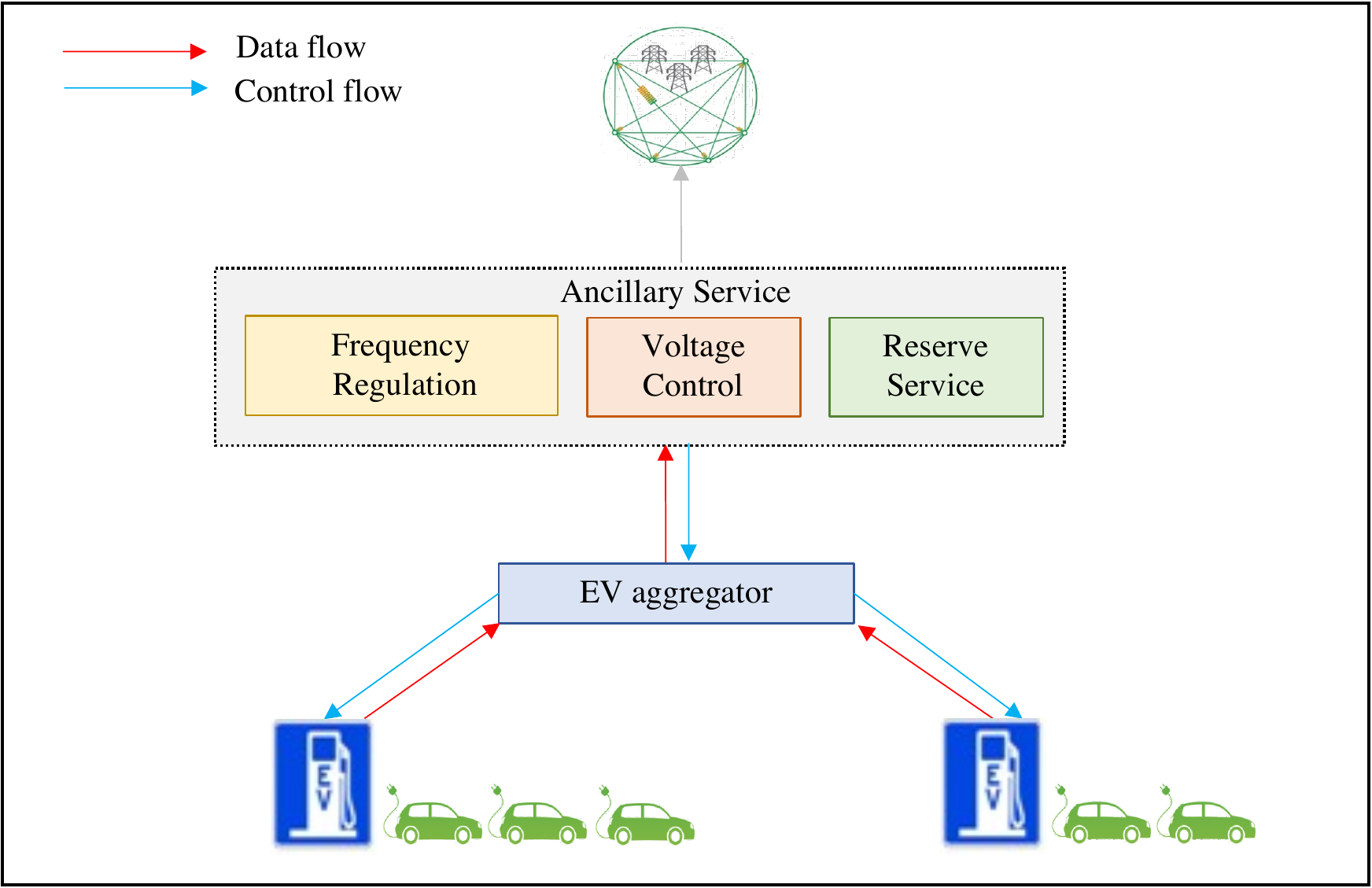}
    \caption{An illustration of EVs providing ancillary services}
    \label{fig:anc}
\end{figure}

Apart from energy trading, EVA can also provide ancillary service to the smart grid, such as  frequency regulation \cite{dong2020online,wang2020electric,oshnoei2021disturbance,pham2021event,saxena2020event}, voltage control \cite{sun2021hierarchical,huang2019day,gao2021voltage,mejia2021coordinated}, and reserve service \cite{bessa2013optimization, hoogvliet2017provision} etc. Ancillary services provide the resources the system operator requires to maintain the instantaneous and continuous balance between power generation and load demand in a reliable manner. Fig.~\ref{fig:anc} gives an illustration of EVs providing ancillary services via EVAs. 

\subsubsection{Frequency regulation}

Frequency regulation is an ancillary service that aims to maintain the frequency of the grid around its nominal value (50 Hz or 60 Hz) by controlling  the frequency variations caused by imbalances between power generation and load demand. Typically, frequency control contains three phases with different timescales. The primary frequency control, also known as droop control is usually triggered within a few seconds. The secondary frequency control, also known as automatic generation control (AGC) is triggered within minutes. The tertiary frequency control, namely economic dispatch is triggered within a few minutes if the frequency deviation event does not correct itself through primary or secondary frequency control mechanisms. Due to the fast response time and high ramp rates of Battery Energy Storage System (BESS), primary frequency control at load-side can be provided by single BESS directly or by multiple small-scale BESSs coordinated by a battery aggregator \cite{ zhu2018optimal}. For example, \cite{zhang2016profit} derived the optimal planning and control strategy for BESSs participating in the primary frequency control regulation market. 

Recently, some studies investigated the potential of EVs to provide frequency regulation services when they are plugged into the grid. Specifically, regulation-down can be done by charging the PEV batteries from the grid, and regulation-up can be achieved by discharging the PEV batteries to the grid. For instance, \cite{dong2020online} proposed an online rolling decoder-dispatch framework for
the frequency management of electrical-grid-electric-vehicle systems.
\cite{wang2020electric} proposed a state-space based EVA modeling and control method for frequency regulation. 
\cite{oshnoei2021disturbance} proposed a control scheme to involve the aggregated EVs in frequency regulation by using a tube-based model predictive control in conjunction with a disturbance observer control.
\cite{pham2021event} considered an event-triggered mechanism
(ETM) for multiple frequency services of electric vehicles (EVs) in smart grids.
\cite{saxena2020event} proposed an event triggered control based switching approach for frequency regulation with EV participation.
\cite{zhang2020joint} proposed a hierarchical system model to jointly optimize power flow routing and V2G scheduling for providing regulation service.

\subsubsection{Voltage control}

Voltage control aims to keep the voltage magnitudes in the smart grid close to the nominal values through injection or digestion of reactive power. Conventionally, voltage control is performed in a centralized manner to determine the day-ahead dispatch of on-load tap changer (OLTC), voltage regulators or capacitor banks, which lack the fast-response capability and are ineffective to mitigate fast voltage violation in real time.  

Recently, some work proposed several control mechanisms that utilize the dispatch of EVs \cite{sun2021hierarchical, huang2019day,gao2021voltage,mejia2021coordinated}.  
\cite{sun2021hierarchical} proposed a three-layer hierarchical voltage control framework to mitigate fast voltage violation problems with the dispatch and control of EVs.
\cite{huang2019day} presented an optimization model to flexibly control available PEV battery charging/discharging power based on three-phase power flow and sensitivity approaches.
\cite{gao2021voltage} proposed a two-stage centralized approach to level the power mismatch between the demand forecast and the real time demand in medium voltage grids by means of fast charging stations.
\cite{mejia2021coordinated} proposed a novel optimal hybrid control framework to improve the voltage profile of highly unbalanced Distribution Grids by coordinating the injection of reactive power from multiple off-board Electrical Vehicles (EVs) chargers.

\subsubsection{Reserve Service}
An operating reserve (spinning reserve, supplemental reserve, replacement reserve) is a power source that can quickly be dispatched to ensure that there is sufficient energy generation to meet load in response to a major generator or transmission outage. Spinning reserves are power sources that are already online, synchronized to grid, and can rapidly increase their power output to meet fast changes in demand. Supplemental reserves can be offline and need not to respond immediately. Replacement reserves are used to restore spinning and supplemental reserves to their pre-contingency status.

There are some work that studied the potential benefits of EVs in providing reserve services \cite{bessa2013optimization, hoogvliet2017provision}. In \cite{bessa2013optimization}, an optimization model and two operational management algorithms were described for supporting the participation of an EVA in the day-ahead energy and manual reserve market sessions. \cite{hoogvliet2017provision} simulated the potential monetary benefit that EV could generate by providing the regulation and reserve power to the Dutch market.
 

 

\section{Case Study}\label{sec:case study}
\begin{figure}
\centering
\includegraphics[width=0.75\linewidth]{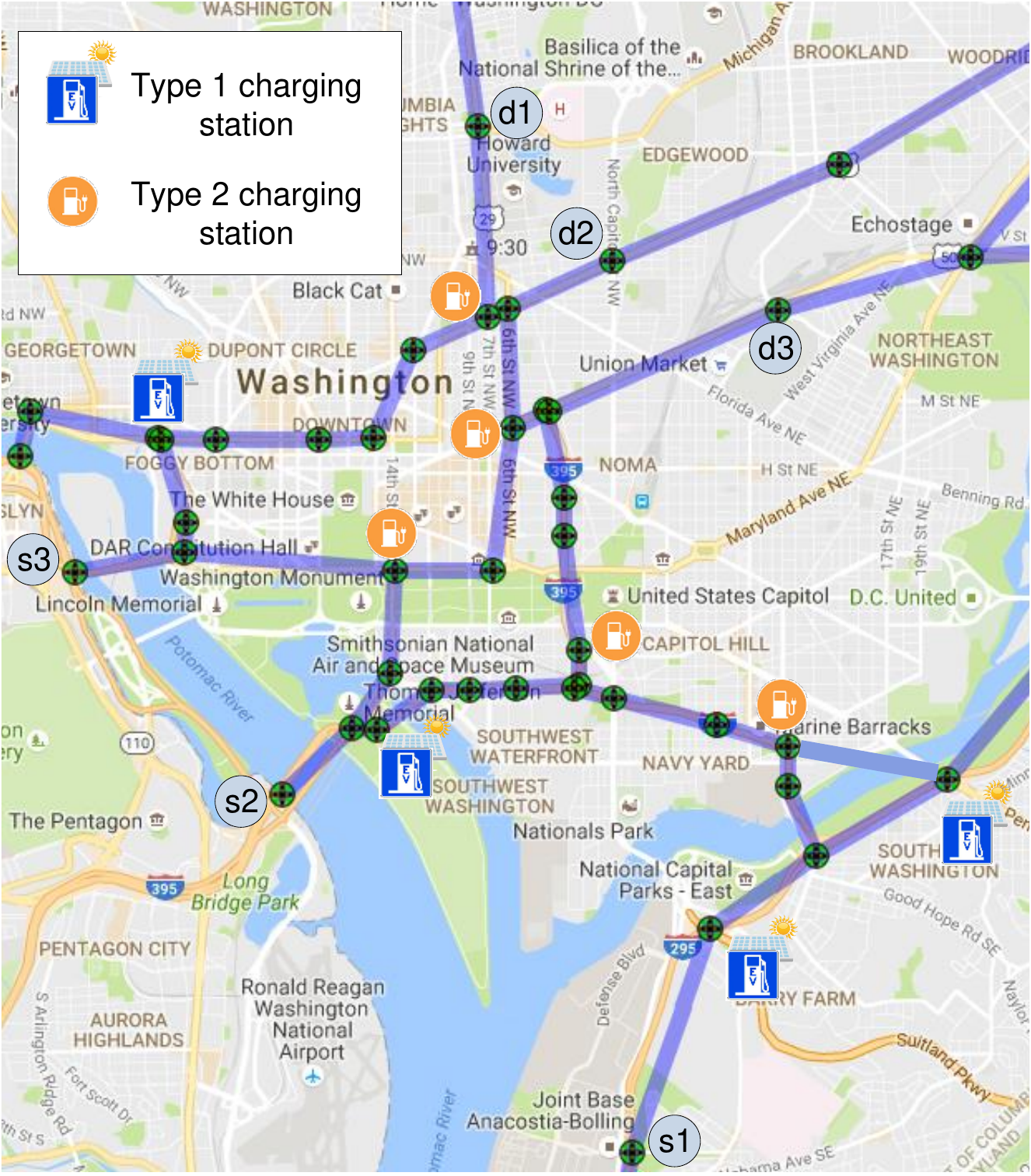}
        \caption{The transportation network used in the case study. }
        \label{fig:map}
\end{figure}

\begin{figure}
    \centering
    \includegraphics[width=0.8\linewidth]{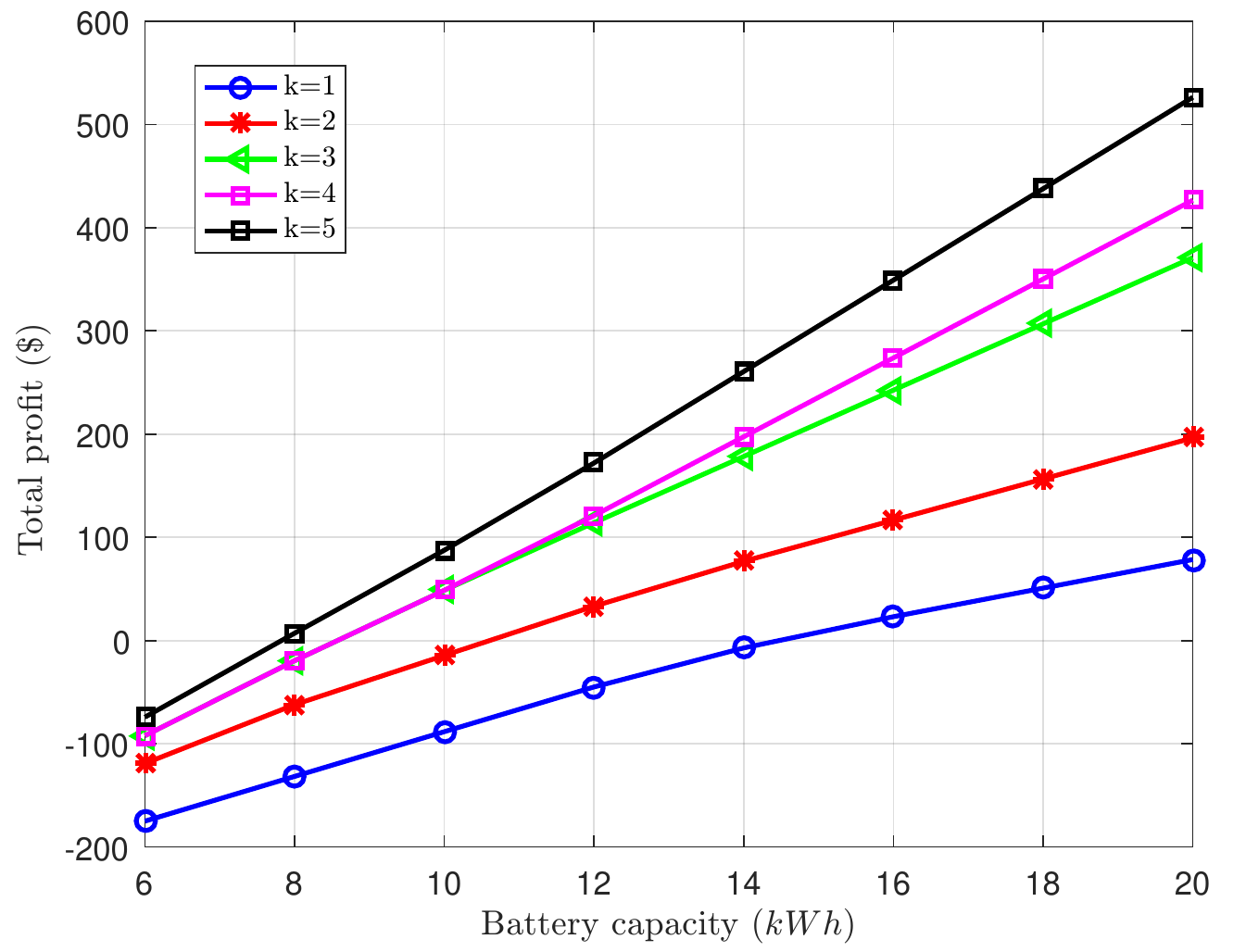}
    \caption{The total profits of all EVs under different routing schemes.}
    \label{fig:totalprofit}
\end{figure}

In this section, we present a case study to illustrate the economic benefit of the joint optimization of routing and charging scheduling of multiple EVs in the IoEV.

We consider the transportation network shown in Fig.~\ref{fig:map}, where the map is plotted based on the data from ``All CHM Plotted Routes, District of Columbia''\footnote{\url{http://courses.teresco.org/chm/viewer/?load=../graphs/dc-all-nomerge.gra}}, which records the geographic coordinates of 61 waypoints and 57 connections between the waypoints.
We adopt the GeographicLib toolbox\footnote{\url{http://www.mathworks.com/matlabcentral/fileexchange/50605-geographiclib}} to calculate the distances of 57 connections.
If the distance between two nodes is no larger than $0.2km$, we add a connection between this pair of nodes.

We consider two types of charging stations deployed in the system,  where type 1 charging stations are powered by renewables and type 2 charging stations are powered by the main grid. 
We assume that type-1 charging stations cannot buy energy from the EVs, as they are installed to transfer the harvested renewable energy to the EVs. In contrast, type 2 charging station can discharge the EVs during the peak hour to reduce the peak load, since a type 2 charging station is typically located in the central of city, which has good connection to the power grid but insufficient space to deploy PV panels. 
In Fig.~\ref{fig:map}, there are four type 1 charging stations, located in suburb areas, and five type 2 charging stations, located in downtown areas.
Type 2 charging stations sell (buy)  the electricity to (from) the EVs at a price of $10 \$/kWh$ ($8 \$/kWh$).
Type 1 charging stations sell renewable energy to the EVs at a very low price of $1 \$/kWh$.
The total amount of renewable energy in each type 1 charging station is set to be $30 kWh$.
Suppose that EV $1, 2$ and $3$ travel from source node $s1, s2$ and $s3$ to destination $d1, d2$ and $d3$, respectively, as shown in Fig.~\ref{fig:map}.
For each EV, the  battery capacity is set to $15 kWh$ and the initial state of charge (SoC) at the source node is set to $0.5$.
The charging efficiency is $0.9$.
Due to the limited capacity of charging station, the route selections of all the EVs are coupled, necessitating a joint optimization across the system.

To reduce the computational complexity and information exchange, we proposed a distributed routing scheme in \cite{tang2019distributed}, where each EV user's selfish behavior to maximize their own profits also leads to the maximum social surplus.
Specifically, the distributed scheme in \cite{tang2019distributed} optimizes the route selection of all the EVs given a set of $k$ shortest paths as candidate paths.
In a special case when $k=1$, the EVs have no choice but to travel through the shortest path from the starting point to the destination, which is equivalent to the conventional distance-based shortest path method. For each given path, the optimal charging scheduling of an EV is to charge its battery as much as possible whenever it encounters a type 1 charging station, and to sell the electricity back to the grid when it encounters a type 2 charging station under the constraint that the remaining energy is sufficient to reach the next type 1 charging station or the destination, whichever is closer.
We plot in Fig.~\ref{fig:totalprofit} the optimal total profits of all EVs under different value of $k$.
It can be observed that the total profit increases when $k$ increases, as increasing k enlarges the solution set of joint route selection and charging scheduling design that aims to maximize the total profit rather than just minimizing the traveling distance. The total profit also grows with the battery capacity. Specifically,
when the battery capacity increases to $20 kWh$, the distributed scheme leads to $447.8491\$$ more profit when $k=5$ compared to when $k=1$, 
which demonstrates the superiority of joint route selection and charging scheduling design over conventional distance-based shortest path method in IoEV networks.

\section{Future Directions} \label{dis}

\subsection{Leveraging data-driven approaches}
\subsubsection{For demand response}
Most existing work on demand response scheme is based on mathematically convenient models that are often too simple to be practical.
Recent success in machine learning holds significant potential in solving this issue by learning from massive real-world data. For instance, EV users may exhibit certain group patterns that can be identified by clustering methods \cite{sun2020classification}. 
The identified charging pattern and user preferences allow us to design more realistic models to handle user heterogeneity. The actual charging profiles of EVs can also help develop more accurate battery models, which enables the refinement on demand-response scheme design. 
In addition, model-free reinforcement learning approaches have shown great success in solving demand response problems, as it can integrate user preference and adapt to the environment.

\subsubsection{For charging and routing optimization}
The EV routing optimization problem in \cite{tang2019distributed} only considers one-stage routing optimization, i.e., one journey with a pair of starting and destination nodes for each EV.  In practice, EVs need to complete multiple journeys. The future journeys are often coupled with the current routing decision and are random in general.
The problem becomes even more challenging when considering the uncertainties from both transportation network and smart grid, e.g., vehicle behaviors, charging habits, time-varying electricity prices, and real-time renewable generations.
Currently, there are several preliminary studies on online and stochastic EV charging scheduling to tackle different types of uncertainties under different types of knowledge of future data \cite{tang2016online}\cite{tang2016model}.
However, most of them suffer from high computational complexity.
A promising solution is to exploit data-driven learning approaches to adapt the decisions to the dynamic environment. For instance, methods like graph neural network can predict the time-varying demand using historical data. Meanwhile, reinforcement learning methods can be designed to learn optimal policy for EV charging scheduling and routing.

\subsubsection{For charging station planning}
Another important issue is to optimize the city-wide charging station/battery swapping placement to
maximize the overall IoEV efficiency. Most existing work on charging station planning focus on proposing mathematical formulation such as mixed-integer linear programming or discrete optimization under a variety of assumptions and validated the formulation using simulation.
Notice that historical EV trajectory data and user behavior data (charging pattern, arrival and departure pattern, etc) are likely to reveal key information of the fine-grained charging demand at different time/location throughout a day. In addition, high-resolution population data such as LandScan and Worldpop can further facilitate the estimation of the long term charging demand. Hence, a data-driven approach is preferred for the charging network design that can better accommodate with the spatially and temporally varying charging demand.


\subsection{Exploit an economic perspective}
\subsubsection{Profit model for providing ancillary service}
The studies so far mainly focused on designing operation and control strategies for EVs to provide ancillary service like frequency regulation and voltage control. However an important aspect is to incentivize EV users to participate in the process in practice. Unlike BESSs, the mobility of EVs brings  the opportunity to respond quickly to the unexpected event of the power grid. In the meantime, a great challenge here is to characterize and model the demand and flexibility in a spatial-temporal context, which should be carefully considered while developing a profit model that can motivate EVs to provide ancillary service.

\subsubsection{Cooperation and competition among different charging station operators}
Most existing work investigated the operation strategy of a single charging station or a charging station operator that coordinates a number of charging stations. In reality, multiple charging station operators owned by different companies often coexist, and new investors may also enter into the market. One interesting direction is to model the cooperation and competition among different charging station operators and design pricing strategies under different perspectives and considerations. In
\cite{zhang2020plug}, the optimal pricing contract and quantity contract for service providers is studied in the duopoly market with Bertrand competition and Cournot competition. Pricing strategies for other scenarios are worth further investigation.

\subsubsection{Data privacy concern}
Efficient real-time charging  scheduling algorithms often require both historic and real-time data from the charging facilities/utilities and EV users.
The data often contains private information, such as location information, travel destination, models of EVs, and individual consumer profiles.
Hence privacy-preserving data collection and processing is a practical and interesting research problem.
For example, efficient incentive schemes can be designed to ensure the truthfulness of the collected data.
To encourage EV users to share the accurate data with the system operator, incentive mechanisms based on, for example, pricing, auction and contract theory, plays an important role. 
In some cases, EV users are not willing to reveal their demand response functions due to the privacy concerns, causing difficulty on pricing scheme design. In this case, a viable approach is to predict the demand functions by learning from the historical data collected by the charging stations.

\subsection{Integration of advanced charging facilities}
Most existing algorithms of joint optimization of EV routing and charging scheduling only considers G2V.
In fact, it has been shown that deployment of V2G and V2V can largely improve the flexibility of energy storage systems and benefit both EV users and  smart grid. For instance, with the implementation of V2G technology, EVs have the incentive to buy cheap electricity from the renewable charging stations and then sell the electricity to the charging stations with heavy load demands.
Accordingly, the routing decisions of the EVs are significantly different when the V2G technology is available. 
In addition, wireless charging facilities can be deployed on the roadway to charge EVs on the move. 
On the other hand, despite its convenience, wireless charging on the road may encourage slow driving speed, as the amount of energy charged to an EV per unit distance is inversely proportional to its driving speed. Thus, the optimization problem needs also take into account the road congestion levels.
This leads to a whole host of new problems that require close coordination of IoEV, smart grid, and transportation systems.

\section{Conclusion} \label{con}
In this article, we introduced the EV charging control problems in two directions: G2V and V2G.  For G2V, we discussed the problem setting and charging control techniques for a single charging station and for charging stations in coupled transportation and power networks. 
For V2G, we illustrated how EVs can perform energy trading in the electricity market and provide ancillary services to the power grid.
Besides, we highlighted some open problems and future research directions of charging scheduling problem for IoEVs.
It is foreseeable that advanced charging technologies for IoEV  will spur new research interests, which finally leads to a highly efficient, reliable, and sustainable smart power grid, intelligent transportation network, and smart city.

\bibliographystyle{ACM-Reference-Format}
\bibliography{ref}


\begin{thebibliography}{55}


\ifx \showCODEN    \undefined \def \showCODEN     #1{\unskip}     \fi
\ifx \showDOI      \undefined \def \showDOI       #1{#1}\fi
\ifx \showISBNx    \undefined \def \showISBNx     #1{\unskip}     \fi
\ifx \showISBNxiii \undefined \def \showISBNxiii  #1{\unskip}     \fi
\ifx \showISSN     \undefined \def \showISSN      #1{\unskip}     \fi
\ifx \showLCCN     \undefined \def \showLCCN      #1{\unskip}     \fi
\ifx \shownote     \undefined \def \shownote      #1{#1}          \fi
\ifx \showarticletitle \undefined \def \showarticletitle #1{#1}   \fi
\ifx \showURL      \undefined \def \showURL       {\relax}        \fi
\providecommand\bibfield[2]{#2}
\providecommand\bibinfo[2]{#2}
\providecommand\natexlab[1]{#1}
\providecommand\showeprint[2][]{arXiv:#2}

\bibitem[\protect\citeauthoryear{Adulyasak and Jaillet}{Adulyasak and
  Jaillet}{2016}]%
        {adulyasak2016models}
\bibfield{author}{\bibinfo{person}{Yossiri Adulyasak} {and}
  \bibinfo{person}{Patrick Jaillet}.} \bibinfo{year}{2016}\natexlab{}.
\newblock \showarticletitle{Models and algorithms for stochastic and robust
  vehicle routing with deadlines}.
\newblock \bibinfo{journal}{\emph{Transportation Science}}
  \bibinfo{volume}{50}, \bibinfo{number}{2} (\bibinfo{year}{2016}),
  \bibinfo{pages}{608--626}.
\newblock


\bibitem[\protect\citeauthoryear{Alizadeh, Wai, Scaglione, Goldsmith, Fan, and
  Javidi}{Alizadeh et~al\mbox{.}}{2014}]%
        {alizadeh2014optimized}
\bibfield{author}{\bibinfo{person}{Mahnoosh Alizadeh}, \bibinfo{person}{Hoi-To
  Wai}, \bibinfo{person}{Anna Scaglione}, \bibinfo{person}{Andrea Goldsmith},
  \bibinfo{person}{Yue~Yue Fan}, {and} \bibinfo{person}{Tara Javidi}.}
  \bibinfo{year}{2014}\natexlab{}.
\newblock \showarticletitle{Optimized path planning for electric vehicle
  routing and charging}. In \bibinfo{booktitle}{\emph{2014 52nd Annual Allerton
  Conference on Communication, Control, and Computing (Allerton)}}.
  \bibinfo{publisher}{IEEE}, \bibinfo{pages}{25--32}.
\newblock


\bibitem[\protect\citeauthoryear{Asrari, Ansari, Khazaei, and Fajri}{Asrari
  et~al\mbox{.}}{2020}]%
        {asrari2019market}
\bibfield{author}{\bibinfo{person}{Arash Asrari}, \bibinfo{person}{Meisam
  Ansari}, \bibinfo{person}{Javad Khazaei}, {and} \bibinfo{person}{Poria
  Fajri}.} \bibinfo{year}{2020}\natexlab{}.
\newblock \showarticletitle{A market framework for decentralized congestion
  management in smart distribution grids considering collaboration among
  electric vehicle aggregators}.
\newblock \bibinfo{journal}{\emph{IEEE Transactions on Smart Grid}}
  \bibinfo{volume}{11}, \bibinfo{number}{2} (\bibinfo{year}{2020}),
  \bibinfo{pages}{1147--1158}.
\newblock


\bibitem[\protect\citeauthoryear{Bessa and Matos}{Bessa and Matos}{2013}]%
        {bessa2013optimization}
\bibfield{author}{\bibinfo{person}{Ricardo~J Bessa} {and}
  \bibinfo{person}{Manuel~A Matos}.} \bibinfo{year}{2013}\natexlab{}.
\newblock \showarticletitle{Optimization models for EV aggregator participation
  in a manual reserve market}.
\newblock \bibinfo{journal}{\emph{IEEE Transactions on Power Systems}}
  \bibinfo{volume}{28}, \bibinfo{number}{3} (\bibinfo{year}{2013}),
  \bibinfo{pages}{3085--3095}.
\newblock


\bibitem[\protect\citeauthoryear{Cao, Song, Kaiwartya, Zhou, Zhuang, Cao, and
  Zhang}{Cao et~al\mbox{.}}{2018}]%
        {cao2018mobile}
\bibfield{author}{\bibinfo{person}{Yue Cao}, \bibinfo{person}{Houbing Song},
  \bibinfo{person}{Omprakash Kaiwartya}, \bibinfo{person}{Bingpeng Zhou},
  \bibinfo{person}{Yuan Zhuang}, \bibinfo{person}{Yang Cao}, {and}
  \bibinfo{person}{Xu Zhang}.} \bibinfo{year}{2018}\natexlab{}.
\newblock \showarticletitle{Mobile edge computing for big-data-enabled electric
  vehicle charging}.
\newblock \bibinfo{journal}{\emph{IEEE Communications Magazine}}
  \bibinfo{volume}{56}, \bibinfo{number}{3} (\bibinfo{year}{2018}),
  \bibinfo{pages}{150--156}.
\newblock


\bibitem[\protect\citeauthoryear{Cerna, Pourakbari-Kasmaei, Romero, and
  Rider}{Cerna et~al\mbox{.}}{2018}]%
        {Cerna2017optimal}
\bibfield{author}{\bibinfo{person}{Fernando~V Cerna}, \bibinfo{person}{Mahdi
  Pourakbari-Kasmaei}, \bibinfo{person}{Ruben~A Romero}, {and}
  \bibinfo{person}{Marcos~J Rider}.} \bibinfo{year}{2018}\natexlab{}.
\newblock \showarticletitle{Optimal delivery scheduling and charging of EVs in
  the navigation of a city map}.
\newblock \bibinfo{journal}{\emph{IEEE Transactions on Smart Grid}}
  \bibinfo{volume}{9}, \bibinfo{number}{5} (\bibinfo{year}{2018}),
  \bibinfo{pages}{4815--4827}.
\newblock


\bibitem[\protect\citeauthoryear{Da~Silva, Nishida, Roijers, and
  Costa}{Da~Silva et~al\mbox{.}}{2019}]%
        {da2019coordination}
\bibfield{author}{\bibinfo{person}{Felipe~Leno Da~Silva},
  \bibinfo{person}{Cyntia~EH Nishida}, \bibinfo{person}{Diederik~M Roijers},
  {and} \bibinfo{person}{Anna H~Reali Costa}.} \bibinfo{year}{2019}\natexlab{}.
\newblock \showarticletitle{Coordination of electric vehicle charging through
  multiagent reinforcement learning}.
\newblock \bibinfo{journal}{\emph{IEEE Transactions on Smart Grid}}
  \bibinfo{volume}{11}, \bibinfo{number}{3} (\bibinfo{year}{2019}),
  \bibinfo{pages}{2347--2356}.
\newblock


\bibitem[\protect\citeauthoryear{Dong, Chu, Morstyn, McCulloch, and Jia}{Dong
  et~al\mbox{.}}{2020}]%
        {dong2020online}
\bibfield{author}{\bibinfo{person}{Chaoyu Dong}, \bibinfo{person}{Ronghe Chu},
  \bibinfo{person}{Thomas Morstyn}, \bibinfo{person}{Malcolm~D McCulloch},
  {and} \bibinfo{person}{Hongjie Jia}.} \bibinfo{year}{2020}\natexlab{}.
\newblock \showarticletitle{Online Rolling Evolutionary Decoder-Dispatch
  Framework for the Secondary Frequency Regulation of Time-Varying
  Electrical-Grid-Electric-Vehicle System}.
\newblock \bibinfo{journal}{\emph{IEEE Transactions on Smart Grid}}
  \bibinfo{volume}{12}, \bibinfo{number}{1} (\bibinfo{year}{2020}),
  \bibinfo{pages}{871--884}.
\newblock


\bibitem[\protect\citeauthoryear{Elghitani and El-Saadany}{Elghitani and
  El-Saadany}{2021}]%
        {elghitani2020efficient}
\bibfield{author}{\bibinfo{person}{Fadi Elghitani} {and}
  \bibinfo{person}{Ehab~F El-Saadany}.} \bibinfo{year}{2021}\natexlab{}.
\newblock \showarticletitle{Efficient assignment of electric vehicles to
  charging stations}.
\newblock \bibinfo{journal}{\emph{IEEE Transactions on Smart Grid}}
  \bibinfo{volume}{12}, \bibinfo{number}{1} (\bibinfo{year}{2021}),
  \bibinfo{pages}{761--773}.
\newblock


\bibitem[\protect\citeauthoryear{Fallah-Mehrjardi, Yaghmaee, and
  Leon-Garcia}{Fallah-Mehrjardi et~al\mbox{.}}{2020}]%
        {fallah2020charge}
\bibfield{author}{\bibinfo{person}{Omid Fallah-Mehrjardi},
  \bibinfo{person}{Mohammad~Hossein Yaghmaee}, {and} \bibinfo{person}{Alberto
  Leon-Garcia}.} \bibinfo{year}{2020}\natexlab{}.
\newblock \showarticletitle{Charge scheduling of electric vehicles in smart
  parking-lot under future demands uncertainty}.
\newblock \bibinfo{journal}{\emph{IEEE Transactions on Smart Grid}}
  \bibinfo{volume}{11}, \bibinfo{number}{6} (\bibinfo{year}{2020}),
  \bibinfo{pages}{4949--4959}.
\newblock


\bibitem[\protect\citeauthoryear{Gao, De~Carne, Andresen, Br{\"u}ske, Pugliese,
  and Liserre}{Gao et~al\mbox{.}}{2021}]%
        {gao2021voltage}
\bibfield{author}{\bibinfo{person}{Xiang Gao}, \bibinfo{person}{Giovanni
  De~Carne}, \bibinfo{person}{Markus Andresen}, \bibinfo{person}{Sebastian
  Br{\"u}ske}, \bibinfo{person}{Sante Pugliese}, {and} \bibinfo{person}{Marco
  Liserre}.} \bibinfo{year}{2021}\natexlab{}.
\newblock \showarticletitle{Voltage-Dependent Load Levelling Approach by means
  of Electric Vehicle Fast Charging Stations}.
\newblock \bibinfo{journal}{\emph{IEEE Transactions on Transportation
  Electrification}} (\bibinfo{year}{2021}).
\newblock


\bibitem[\protect\citeauthoryear{Hajebrahimi, Kamwa, Abdelaziz, and
  Moeini}{Hajebrahimi et~al\mbox{.}}{2020}]%
        {hajebrahimi2020scenario}
\bibfield{author}{\bibinfo{person}{Ali Hajebrahimi}, \bibinfo{person}{Innocent
  Kamwa}, \bibinfo{person}{Morad Mohamed~Abdelmageed Abdelaziz}, {and}
  \bibinfo{person}{Ali Moeini}.} \bibinfo{year}{2020}\natexlab{}.
\newblock \showarticletitle{Scenario-wise distributionally robust optimization
  for collaborative intermittent resources and electric vehicle aggregator
  bidding strategy}.
\newblock \bibinfo{journal}{\emph{IEEE Transactions on Power Systems}}
  \bibinfo{volume}{35}, \bibinfo{number}{5} (\bibinfo{year}{2020}),
  \bibinfo{pages}{3706--3718}.
\newblock


\bibitem[\protect\citeauthoryear{Han, Lee, and Park}{Han et~al\mbox{.}}{2020}]%
        {han2020optimal}
\bibfield{author}{\bibinfo{person}{Sini Han}, \bibinfo{person}{Duehee Lee},
  {and} \bibinfo{person}{Jong-Bae Park}.} \bibinfo{year}{2020}\natexlab{}.
\newblock \showarticletitle{Optimal Bidding and Operation Strategies for EV
  Aggegators by Regrouping Aggregated EV Batteries}.
\newblock \bibinfo{journal}{\emph{IEEE Transactions on Smart Grid}}
  \bibinfo{volume}{11}, \bibinfo{number}{6} (\bibinfo{year}{2020}),
  \bibinfo{pages}{4928--4937}.
\newblock


\bibitem[\protect\citeauthoryear{Hoogvliet, Litjens, and Van~Sark}{Hoogvliet
  et~al\mbox{.}}{2017}]%
        {hoogvliet2017provision}
\bibfield{author}{\bibinfo{person}{TW Hoogvliet}, \bibinfo{person}{GBMA
  Litjens}, {and} \bibinfo{person}{WGJHM Van~Sark}.}
  \bibinfo{year}{2017}\natexlab{}.
\newblock \showarticletitle{Provision of regulating-and reserve power by
  electric vehicle owners in the Dutch market}.
\newblock \bibinfo{journal}{\emph{Applied Energy}}  \bibinfo{volume}{190}
  (\bibinfo{year}{2017}), \bibinfo{pages}{1008--1019}.
\newblock


\bibitem[\protect\citeauthoryear{Huang}{Huang}{2019}]%
        {huang2019day}
\bibfield{author}{\bibinfo{person}{Yulong Huang}.}
  \bibinfo{year}{2019}\natexlab{}.
\newblock \showarticletitle{Day-ahead optimal control of PEV battery storage
  devices taking into account the voltage regulation of the residential power
  grid}.
\newblock \bibinfo{journal}{\emph{IEEE Transactions on Power Systems}}
  \bibinfo{volume}{34}, \bibinfo{number}{6} (\bibinfo{year}{2019}),
  \bibinfo{pages}{4154--4167}.
\newblock


\bibitem[\protect\citeauthoryear{Kempton and Tomi{\'c}}{Kempton and
  Tomi{\'c}}{2005}]%
        {kempton2005vehicle}
\bibfield{author}{\bibinfo{person}{Willett Kempton} {and}
  \bibinfo{person}{Jasna Tomi{\'c}}.} \bibinfo{year}{2005}\natexlab{}.
\newblock \showarticletitle{Vehicle-to-grid power implementation: From
  stabilizing the grid to supporting large-scale renewable energy}.
\newblock \bibinfo{journal}{\emph{Journal of power sources}}
  \bibinfo{volume}{144}, \bibinfo{number}{1} (\bibinfo{year}{2005}),
  \bibinfo{pages}{280--294}.
\newblock


\bibitem[\protect\citeauthoryear{Lam and Li}{Lam and Li}{2018}]%
        {lam2018opportunistic}
\bibfield{author}{\bibinfo{person}{Albert~YS Lam} {and}
  \bibinfo{person}{Victor~OK Li}.} \bibinfo{year}{2018}\natexlab{}.
\newblock \showarticletitle{Opportunistic routing for vehicular energy
  network}.
\newblock \bibinfo{journal}{\emph{IEEE Internet of Things Journal}}
  \bibinfo{volume}{5}, \bibinfo{number}{2} (\bibinfo{year}{2018}),
  \bibinfo{pages}{533--545}.
\newblock


\bibitem[\protect\citeauthoryear{Lee, Lee, Lee, Jin, Lee, Low, Chang, Ortega,
  and Low}{Lee et~al\mbox{.}}{2020}]%
        {lee2020adaptive}
\bibfield{author}{\bibinfo{person}{Zachary~J Lee}, \bibinfo{person}{George
  Lee}, \bibinfo{person}{Ted Lee}, \bibinfo{person}{Cheng Jin},
  \bibinfo{person}{Rand Lee}, \bibinfo{person}{Zhi Low},
  \bibinfo{person}{Daniel Chang}, \bibinfo{person}{Christine Ortega}, {and}
  \bibinfo{person}{Steven~H Low}.} \bibinfo{year}{2020}\natexlab{}.
\newblock \showarticletitle{Adaptive Charging Networks: A Framework for Smart
  Electric Vehicle Charging}.
\newblock \bibinfo{journal}{\emph{arXiv preprint arXiv:2012.02636}}
  (\bibinfo{year}{2020}).
\newblock


\bibitem[\protect\citeauthoryear{Li, Wan, and He}{Li et~al\mbox{.}}{2020}]%
        {li2019constrained}
\bibfield{author}{\bibinfo{person}{Hepeng Li}, \bibinfo{person}{Zhiqiang Wan},
  {and} \bibinfo{person}{Haibo He}.} \bibinfo{year}{2020}\natexlab{}.
\newblock \showarticletitle{Constrained EV charging scheduling based on safe
  deep reinforcement learning}.
\newblock \bibinfo{journal}{\emph{IEEE Transactions on Smart Grid}}
  \bibinfo{volume}{11}, \bibinfo{number}{3} (\bibinfo{year}{2020}),
  \bibinfo{pages}{2427--2439}.
\newblock


\bibitem[\protect\citeauthoryear{Liu, Lin, Huang, Zhou, Li, and Rehtanz}{Liu
  et~al\mbox{.}}{2020}]%
        {liu2020optimal}
\bibfield{author}{\bibinfo{person}{Jiayan Liu}, \bibinfo{person}{Gang Lin},
  \bibinfo{person}{Sunhua Huang}, \bibinfo{person}{Yang Zhou},
  \bibinfo{person}{Yong Li}, {and} \bibinfo{person}{Christian Rehtanz}.}
  \bibinfo{year}{2020}\natexlab{}.
\newblock \showarticletitle{Optimal Logistics EV Charging Scheduling by
  Considering the Limited Number of Chargers}.
\newblock \bibinfo{journal}{\emph{IEEE Transactions on Transportation
  Electrification}} (\bibinfo{year}{2020}).
\newblock


\bibitem[\protect\citeauthoryear{Liu, Chen, Hou, and Yang}{Liu
  et~al\mbox{.}}{2021}]%
        {liu2021optimal}
\bibfield{author}{\bibinfo{person}{Wenjie Liu}, \bibinfo{person}{Shibo Chen},
  \bibinfo{person}{Yunhe Hou}, {and} \bibinfo{person}{Zaiyue Yang}.}
  \bibinfo{year}{2021}\natexlab{}.
\newblock \showarticletitle{Optimal Reserve Management of Electric Vehicle
  Aggregator: Discrete Bilevel Optimization Model and Exact Algorithm}.
\newblock \bibinfo{journal}{\emph{IEEE Transactions on Smart Grid}}
  (\bibinfo{year}{2021}).
\newblock


\bibitem[\protect\citeauthoryear{Long, Jia, Wang, and Yang}{Long
  et~al\mbox{.}}{2021}]%
        {long2021efficient}
\bibfield{author}{\bibinfo{person}{Teng Long}, \bibinfo{person}{Qing-Shan Jia},
  \bibinfo{person}{Gongming Wang}, {and} \bibinfo{person}{Yu Yang}.}
  \bibinfo{year}{2021}\natexlab{}.
\newblock \showarticletitle{Efficient Real-Time EV Charging Scheduling via
  Ordinal Optimization}.
\newblock \bibinfo{journal}{\emph{IEEE Transactions on Smart Grid}}
  (\bibinfo{year}{2021}).
\newblock


\bibitem[\protect\citeauthoryear{Madahi, Nafisi, Abyaneh, and Marzband}{Madahi
  et~al\mbox{.}}{2021}]%
        {madahi2020co}
\bibfield{author}{\bibinfo{person}{Seyed Soroush~Karimi Madahi},
  \bibinfo{person}{Hamed Nafisi}, \bibinfo{person}{Hossein~Askarian Abyaneh},
  {and} \bibinfo{person}{Mousa Marzband}.} \bibinfo{year}{2021}\natexlab{}.
\newblock \showarticletitle{Co-Optimization of Energy Losses and Transformer
  Operating Costs Based on Smart Charging Algorithm for Plug-in Electric
  Vehicle Parking Lots}.
\newblock \bibinfo{journal}{\emph{IEEE Transactions on Transportation
  Electrification}} \bibinfo{volume}{7}, \bibinfo{number}{2}
  (\bibinfo{year}{2021}), \bibinfo{pages}{527--541}.
\newblock


\bibitem[\protect\citeauthoryear{Mejia-Ruiz, C{\'a}rdenas-Javier, Paternina,
  Rodr{\'\i}guez-Rodr{\'\i}guez, Ramirez, and Zamora-Mendez}{Mejia-Ruiz
  et~al\mbox{.}}{2021}]%
        {mejia2021coordinated}
\bibfield{author}{\bibinfo{person}{Gabriel~E Mejia-Ruiz},
  \bibinfo{person}{Romel C{\'a}rdenas-Javier}, \bibinfo{person}{Mario R~Arrieta
  Paternina}, \bibinfo{person}{Juan~R Rodr{\'\i}guez-Rodr{\'\i}guez},
  \bibinfo{person}{Juan~M Ramirez}, {and} \bibinfo{person}{Alejandro
  Zamora-Mendez}.} \bibinfo{year}{2021}\natexlab{}.
\newblock \showarticletitle{Coordinated Optimal Volt/Var Control for
  Distribution Networks via D-PMUs and EV Chargers by Exploiting the
  Eigensystem Realization}.
\newblock \bibinfo{journal}{\emph{IEEE Transactions on Smart Grid}}
  \bibinfo{volume}{12}, \bibinfo{number}{3} (\bibinfo{year}{2021}),
  \bibinfo{pages}{2425--2438}.
\newblock


\bibitem[\protect\citeauthoryear{Moghaddam, Ahmad, Habibi, and
  Masoum}{Moghaddam et~al\mbox{.}}{2019}]%
        {moghaddam2019coordinated}
\bibfield{author}{\bibinfo{person}{Zeinab Moghaddam}, \bibinfo{person}{Iftekhar
  Ahmad}, \bibinfo{person}{Daryoush Habibi}, {and} \bibinfo{person}{Mohammad~AS
  Masoum}.} \bibinfo{year}{2019}\natexlab{}.
\newblock \showarticletitle{A coordinated dynamic pricing model for electric
  vehicle charging stations}.
\newblock \bibinfo{journal}{\emph{IEEE Transactions on Transportation
  Electrification}} \bibinfo{volume}{5}, \bibinfo{number}{1}
  (\bibinfo{year}{2019}), \bibinfo{pages}{226--238}.
\newblock


\bibitem[\protect\citeauthoryear{Moghaddass, Mohammed, Skordilis, and
  Asfour}{Moghaddass et~al\mbox{.}}{2019}]%
        {moghaddass2019smart}
\bibfield{author}{\bibinfo{person}{Ramin Moghaddass}, \bibinfo{person}{Osama~A
  Mohammed}, \bibinfo{person}{Erotokritos Skordilis}, {and}
  \bibinfo{person}{Shihab Asfour}.} \bibinfo{year}{2019}\natexlab{}.
\newblock \showarticletitle{Smart control of fleets of electric vehicles in
  smart and connected communities}.
\newblock \bibinfo{journal}{\emph{IEEE Transactions on Smart Grid}}
  \bibinfo{volume}{10}, \bibinfo{number}{6} (\bibinfo{year}{2019}),
  \bibinfo{pages}{6883--6897}.
\newblock


\bibitem[\protect\citeauthoryear{Moradipari and Alizadeh}{Moradipari and
  Alizadeh}{2019}]%
        {moradipari2019pricing}
\bibfield{author}{\bibinfo{person}{Ahmadreza Moradipari} {and}
  \bibinfo{person}{Mahnoosh Alizadeh}.} \bibinfo{year}{2019}\natexlab{}.
\newblock \showarticletitle{Pricing and routing mechanisms for differentiated
  services in an electric vehicle public charging station network}.
\newblock \bibinfo{journal}{\emph{IEEE Transactions on Smart Grid}}
  \bibinfo{volume}{11}, \bibinfo{number}{2} (\bibinfo{year}{2019}),
  \bibinfo{pages}{1489--1499}.
\newblock


\bibitem[\protect\citeauthoryear{Muratori}{Muratori}{2018}]%
        {muratori2018impact}
\bibfield{author}{\bibinfo{person}{Matteo Muratori}.}
  \bibinfo{year}{2018}\natexlab{}.
\newblock \showarticletitle{Impact of uncoordinated plug-in electric vehicle
  charging on residential power demand}.
\newblock \bibinfo{journal}{\emph{Nature Energy}} \bibinfo{volume}{3},
  \bibinfo{number}{3} (\bibinfo{year}{2018}), \bibinfo{pages}{193--201}.
\newblock


\bibitem[\protect\citeauthoryear{Oshnoei, Kheradmandi, Muyeen, and
  Hatziargyriou}{Oshnoei et~al\mbox{.}}{2021}]%
        {oshnoei2021disturbance}
\bibfield{author}{\bibinfo{person}{Arman Oshnoei}, \bibinfo{person}{Morteza
  Kheradmandi}, \bibinfo{person}{SM Muyeen}, {and} \bibinfo{person}{Nikos~D
  Hatziargyriou}.} \bibinfo{year}{2021}\natexlab{}.
\newblock \showarticletitle{Disturbance Observer and Tube-based Model
  Predictive Controlled Electric Vehicles for Frequency Regulation of an
  Isolated Power Grid}.
\newblock \bibinfo{journal}{\emph{IEEE Transactions on Smart Grid}}
  (\bibinfo{year}{2021}).
\newblock


\bibitem[\protect\citeauthoryear{Pham, Amanullah, and Trinh}{Pham
  et~al\mbox{.}}{2021}]%
        {pham2021event}
\bibfield{author}{\bibinfo{person}{Ngoc~Thanh Pham}, \bibinfo{person}{Oo
  Amanullah}, {and} \bibinfo{person}{Hieu Trinh}.}
  \bibinfo{year}{2021}\natexlab{}.
\newblock \showarticletitle{Event-Triggered Mechanism for Multiple Frequency
  Services of Electric Vehicles in Smart Grids}.
\newblock \bibinfo{journal}{\emph{IEEE Transactions on Power Systems}}
  (\bibinfo{year}{2021}).
\newblock


\bibitem[\protect\citeauthoryear{Porras, Fern{\'a}ndez-Blanco, Morales, and
  Pineda}{Porras et~al\mbox{.}}{2020}]%
        {porras2020efficient}
\bibfield{author}{\bibinfo{person}{{\'A}lvaro Porras}, \bibinfo{person}{Ricardo
  Fern{\'a}ndez-Blanco}, \bibinfo{person}{Juan~M Morales}, {and}
  \bibinfo{person}{Salvador Pineda}.} \bibinfo{year}{2020}\natexlab{}.
\newblock \showarticletitle{An Efficient Robust Approach to the Day-ahead
  Operation of an Aggregator of Electric Vehicles}.
\newblock \bibinfo{journal}{\emph{IEEE Transactions on Smart Grid}}
  \bibinfo{volume}{11}, \bibinfo{number}{6} (\bibinfo{year}{2020}),
  \bibinfo{pages}{4960--4970}.
\newblock


\bibitem[\protect\citeauthoryear{Qian, Shao, Li, Wang, and Shahidehpour}{Qian
  et~al\mbox{.}}{2020}]%
        {qian2020enhanced}
\bibfield{author}{\bibinfo{person}{Tao Qian}, \bibinfo{person}{Chengcheng
  Shao}, \bibinfo{person}{Xuliang Li}, \bibinfo{person}{Xiuli Wang}, {and}
  \bibinfo{person}{Mohammad Shahidehpour}.} \bibinfo{year}{2020}\natexlab{}.
\newblock \showarticletitle{Enhanced coordinated operations of electric power
  and transportation networks via EV charging services}.
\newblock \bibinfo{journal}{\emph{IEEE Transactions on Smart Grid}}
  \bibinfo{volume}{11}, \bibinfo{number}{4} (\bibinfo{year}{2020}),
  \bibinfo{pages}{3019--3030}.
\newblock


\bibitem[\protect\citeauthoryear{Sadreddini, Guner, and Erdinc}{Sadreddini
  et~al\mbox{.}}{2021}]%
        {sadreddini2021design}
\bibfield{author}{\bibinfo{person}{Zhaleh Sadreddini}, \bibinfo{person}{Sitki
  Guner}, {and} \bibinfo{person}{Ozan Erdinc}.}
  \bibinfo{year}{2021}\natexlab{}.
\newblock \showarticletitle{Design of a decision-based multi-criteria
  reservation system for the EV parking lot}.
\newblock \bibinfo{journal}{\emph{IEEE Transactions on Transportation
  Electrification}} (\bibinfo{year}{2021}).
\newblock


\bibitem[\protect\citeauthoryear{Saxena and Fridman}{Saxena and
  Fridman}{2020}]%
        {saxena2020event}
\bibfield{author}{\bibinfo{person}{Sahaj Saxena} {and} \bibinfo{person}{Emilia
  Fridman}.} \bibinfo{year}{2020}\natexlab{}.
\newblock \showarticletitle{Event-Triggered load frequency control via
  switching approach}.
\newblock \bibinfo{journal}{\emph{IEEE Transactions on Power Systems}}
  \bibinfo{volume}{35}, \bibinfo{number}{6} (\bibinfo{year}{2020}),
  \bibinfo{pages}{4484--4494}.
\newblock


\bibitem[\protect\citeauthoryear{{\v{S}}epetanc and
  Pand{\v{z}}i{\'c}}{{\v{S}}epetanc and Pand{\v{z}}i{\'c}}{2021}]%
        {vsepetanc2021cluster}
\bibfield{author}{\bibinfo{person}{Karlo {\v{S}}epetanc} {and}
  \bibinfo{person}{H Pand{\v{z}}i{\'c}}.} \bibinfo{year}{2021}\natexlab{}.
\newblock \showarticletitle{A Cluster-Based Model for Charging a Single-Depot
  Fleet of Electric Vehicles}.
\newblock \bibinfo{journal}{\emph{IEEE Transactions on Smart Grid}}
  \bibinfo{volume}{12}, \bibinfo{number}{4} (\bibinfo{year}{2021}),
  \bibinfo{pages}{3339--3352}.
\newblock


\bibitem[\protect\citeauthoryear{Sheng, Tian, and Leung}{Sheng
  et~al\mbox{.}}{2018}]%
        {sheng2018toward}
\bibfield{author}{\bibinfo{person}{Zhengguo Sheng}, \bibinfo{person}{Daxin
  Tian}, {and} \bibinfo{person}{Victor~CM Leung}.}
  \bibinfo{year}{2018}\natexlab{}.
\newblock \showarticletitle{Toward an energy and resource efficient internet of
  things: A design principle combining computation, communications, and
  protocols}.
\newblock \bibinfo{journal}{\emph{IEEE Communications Magazine}}
  \bibinfo{volume}{56}, \bibinfo{number}{7} (\bibinfo{year}{2018}),
  \bibinfo{pages}{89--95}.
\newblock


\bibitem[\protect\citeauthoryear{Sun, Li, Low, and Tsang}{Sun
  et~al\mbox{.}}{2020b}]%
        {sun2020orc}
\bibfield{author}{\bibinfo{person}{Bo Sun}, \bibinfo{person}{Tongxin Li},
  \bibinfo{person}{Steven~H Low}, {and} \bibinfo{person}{Danny~HK Tsang}.}
  \bibinfo{year}{2020}\natexlab{b}.
\newblock \showarticletitle{ORC: An Online Competitive Algorithm for
  Recommendation and Charging Schedule in Electric Vehicle Charging Network}.
  In \bibinfo{booktitle}{\emph{Proceedings of the Eleventh ACM International
  Conference on Future Energy Systems}}. \bibinfo{pages}{144--155}.
\newblock


\bibitem[\protect\citeauthoryear{Sun, Li, Low, and Li}{Sun
  et~al\mbox{.}}{2020a}]%
        {sun2020classification}
\bibfield{author}{\bibinfo{person}{Chenxi Sun}, \bibinfo{person}{Tongxin Li},
  \bibinfo{person}{Steven~H Low}, {and} \bibinfo{person}{Victor~OK Li}.}
  \bibinfo{year}{2020}\natexlab{a}.
\newblock \showarticletitle{Classification of electric vehicle charging time
  series with selective clustering}.
\newblock \bibinfo{journal}{\emph{Electric Power Systems Research}}
  \bibinfo{volume}{189} (\bibinfo{year}{2020}), \bibinfo{pages}{106695}.
\newblock


\bibitem[\protect\citeauthoryear{Sun and Qiu}{Sun and Qiu}{2021}]%
        {sun2021hierarchical}
\bibfield{author}{\bibinfo{person}{Xianzhuo Sun} {and} \bibinfo{person}{Jing
  Qiu}.} \bibinfo{year}{2021}\natexlab{}.
\newblock \showarticletitle{Hierarchical Voltage Control Strategy in
  Distribution Networks Considering Customized Charging Navigation of Electric
  Vehicles}.
\newblock \bibinfo{journal}{\emph{IEEE Transactions on Smart Grid}}
  (\bibinfo{year}{2021}).
\newblock


\bibitem[\protect\citeauthoryear{Tang, Bi, and Zhang}{Tang
  et~al\mbox{.}}{2014}]%
        {tang2014online}
\bibfield{author}{\bibinfo{person}{Wanrong Tang}, \bibinfo{person}{Suzhi Bi},
  {and} \bibinfo{person}{Ying~Jun Zhang}.} \bibinfo{year}{2014}\natexlab{}.
\newblock \showarticletitle{Online coordinated charging decision algorithm for
  electric vehicles without future information}.
\newblock \bibinfo{journal}{\emph{IEEE Transactions on Smart Grid}}
  \bibinfo{volume}{5}, \bibinfo{number}{6} (\bibinfo{year}{2014}),
  \bibinfo{pages}{2810--2824}.
\newblock


\bibitem[\protect\citeauthoryear{Tang, Bi, and Zhang}{Tang
  et~al\mbox{.}}{2016}]%
        {tang2016online}
\bibfield{author}{\bibinfo{person}{Wanrong Tang}, \bibinfo{person}{Suzhi Bi},
  {and} \bibinfo{person}{Ying~Jun Zhang}.} \bibinfo{year}{2016}\natexlab{}.
\newblock \showarticletitle{Online charging scheduling algorithms of electric
  vehicles in smart grid: An overview}.
\newblock \bibinfo{journal}{\emph{IEEE communications Magazine}}
  \bibinfo{volume}{54}, \bibinfo{number}{12} (\bibinfo{year}{2016}),
  \bibinfo{pages}{76--83}.
\newblock


\bibitem[\protect\citeauthoryear{Tang and Zhang}{Tang and Zhang}{2017}]%
        {tang2016model}
\bibfield{author}{\bibinfo{person}{Wanrong Tang} {and}
  \bibinfo{person}{Ying~Jun Zhang}.} \bibinfo{year}{2017}\natexlab{}.
\newblock \showarticletitle{A model predictive control approach for
  low-complexity electric vehicle charging scheduling: Optimality and
  scalability}.
\newblock \bibinfo{journal}{\emph{IEEE transactions on power systems}}
  \bibinfo{volume}{32}, \bibinfo{number}{2} (\bibinfo{year}{2017}),
  \bibinfo{pages}{1050--1063}.
\newblock


\bibitem[\protect\citeauthoryear{Tang, Bi, and Zhang}{Tang
  et~al\mbox{.}}{2019}]%
        {tang2019distributed}
\bibfield{author}{\bibinfo{person}{Xiaoying Tang}, \bibinfo{person}{Suzhi Bi},
  {and} \bibinfo{person}{Ying-Jun~Angela Zhang}.}
  \bibinfo{year}{2019}\natexlab{}.
\newblock \showarticletitle{Distributed routing and charging scheduling
  optimization for internet of electric vehicles}.
\newblock \bibinfo{journal}{\emph{IEEE Internet of Things Journal}}
  \bibinfo{volume}{6}, \bibinfo{number}{1} (\bibinfo{year}{2019}),
  \bibinfo{pages}{136--148}.
\newblock


\bibitem[\protect\citeauthoryear{Wang, Dehghanian, and Zhao}{Wang
  et~al\mbox{.}}{2020a}]%
        {wang2019chance}
\bibfield{author}{\bibinfo{person}{Bo Wang}, \bibinfo{person}{Payman
  Dehghanian}, {and} \bibinfo{person}{Dongbo Zhao}.}
  \bibinfo{year}{2020}\natexlab{a}.
\newblock \showarticletitle{Chance-constrained energy management system for
  power grids with high proliferation of renewables and electric vehicles}.
\newblock \bibinfo{journal}{\emph{IEEE Transactions on Smart Grid}}
  \bibinfo{volume}{11}, \bibinfo{number}{3} (\bibinfo{year}{2020}),
  \bibinfo{pages}{2324--2336}.
\newblock


\bibitem[\protect\citeauthoryear{Wang, Mu, Shi, Jia, and Li}{Wang
  et~al\mbox{.}}{2020b}]%
        {wang2020electric}
\bibfield{author}{\bibinfo{person}{Mingshen Wang}, \bibinfo{person}{Yunfei Mu},
  \bibinfo{person}{Qingxin Shi}, \bibinfo{person}{Hongjie Jia}, {and}
  \bibinfo{person}{Fangxing Li}.} \bibinfo{year}{2020}\natexlab{b}.
\newblock \showarticletitle{Electric vehicle aggregator modeling and control
  for frequency regulation considering progressive state recovery}.
\newblock \bibinfo{journal}{\emph{IEEE Transactions on Smart Grid}}
  \bibinfo{volume}{11}, \bibinfo{number}{5} (\bibinfo{year}{2020}),
  \bibinfo{pages}{4176--4189}.
\newblock


\bibitem[\protect\citeauthoryear{Wang, Bi, and Zhang}{Wang
  et~al\mbox{.}}{2021}]%
        {wang2021rein}
\bibfield{author}{\bibinfo{person}{Shuoyao Wang}, \bibinfo{person}{Suzhi Bi},
  {and} \bibinfo{person}{Yingjun~Angela Zhang}.}
  \bibinfo{year}{2021}\natexlab{}.
\newblock \showarticletitle{Reinforcement learning for real-time pricing and
  scheduling control in EV charging stations}.
\newblock \bibinfo{journal}{\emph{IEEE Transactions on Industrial Informatics}}
  \bibinfo{volume}{17}, \bibinfo{number}{2} (\bibinfo{year}{2021}),
  \bibinfo{pages}{849--859}.
\newblock


\bibitem[\protect\citeauthoryear{Wang, Bi, Zhang, and Huang}{Wang
  et~al\mbox{.}}{2018}]%
        {wang2018electrical}
\bibfield{author}{\bibinfo{person}{Shuoyao Wang}, \bibinfo{person}{Suzhi Bi},
  \bibinfo{person}{Ying-Jun~Angela Zhang}, {and} \bibinfo{person}{Jianwei
  Huang}.} \bibinfo{year}{2018}\natexlab{}.
\newblock \showarticletitle{Electrical vehicle charging station profit
  maximization: Admission, pricing, and online scheduling}.
\newblock \bibinfo{journal}{\emph{IEEE Transactions on Sustainable Energy}}
  \bibinfo{volume}{9}, \bibinfo{number}{4} (\bibinfo{year}{2018}),
  \bibinfo{pages}{1722--1731}.
\newblock


\bibitem[\protect\citeauthoryear{Wang and Thompson}{Wang and Thompson}{2019}]%
        {wang2018two}
\bibfield{author}{\bibinfo{person}{Yuchang Wang} {and} \bibinfo{person}{John~S
  Thompson}.} \bibinfo{year}{2019}\natexlab{}.
\newblock \showarticletitle{Two-stage admission and scheduling mechanism for
  electric vehicle charging}.
\newblock \bibinfo{journal}{\emph{IEEE Transactions on Smart Grid}}
  \bibinfo{volume}{10}, \bibinfo{number}{3} (\bibinfo{year}{2019}),
  \bibinfo{pages}{2650--2660}.
\newblock


\bibitem[\protect\citeauthoryear{Zeballos, Ferragut, and Paganini}{Zeballos
  et~al\mbox{.}}{2019}]%
        {zeballos2019proportional}
\bibfield{author}{\bibinfo{person}{Martin Zeballos}, \bibinfo{person}{Andres
  Ferragut}, {and} \bibinfo{person}{Fernando Paganini}.}
  \bibinfo{year}{2019}\natexlab{}.
\newblock \showarticletitle{Proportional fairness for EV charging in overload}.
\newblock \bibinfo{journal}{\emph{IEEE Transactions on Smart Grid}}
  \bibinfo{volume}{10}, \bibinfo{number}{6} (\bibinfo{year}{2019}),
  \bibinfo{pages}{6792--6801}.
\newblock


\bibitem[\protect\citeauthoryear{Zeng, Bae, Travacca, and Moura}{Zeng
  et~al\mbox{.}}{2021}]%
        {zeng2021inducing}
\bibfield{author}{\bibinfo{person}{Teng Zeng}, \bibinfo{person}{Sangjae Bae},
  \bibinfo{person}{Bertrand Travacca}, {and} \bibinfo{person}{Scott Moura}.}
  \bibinfo{year}{2021}\natexlab{}.
\newblock \showarticletitle{Inducing Human Behavior to Maximize Operation
  Performance at PEV Charging Station}.
\newblock \bibinfo{journal}{\emph{IEEE Transactions on Smart Grid}}
  \bibinfo{volume}{12}, \bibinfo{number}{4} (\bibinfo{year}{2021}),
  \bibinfo{pages}{3353--3363}.
\newblock


\bibitem[\protect\citeauthoryear{Zhang, Hu, and Song}{Zhang
  et~al\mbox{.}}{2020}]%
        {zhang2020power}
\bibfield{author}{\bibinfo{person}{Hongcai Zhang}, \bibinfo{person}{Zechun Hu},
  {and} \bibinfo{person}{Yonghua Song}.} \bibinfo{year}{2020}\natexlab{}.
\newblock \showarticletitle{Power and transport nexus: Routing electric
  vehicles to promote renewable power integration}.
\newblock \bibinfo{journal}{\emph{IEEE Transactions on Smart Grid}}
  \bibinfo{volume}{11}, \bibinfo{number}{4} (\bibinfo{year}{2020}),
  \bibinfo{pages}{3291--3301}.
\newblock


\bibitem[\protect\citeauthoryear{Zhang and Leung}{Zhang and Leung}{2020}]%
        {zhang2020joint}
\bibfield{author}{\bibinfo{person}{Shiyao Zhang} {and}
  \bibinfo{person}{Ka-Cheong Leung}.} \bibinfo{year}{2020}\natexlab{}.
\newblock \showarticletitle{Joint optimal power flow routing and
  vehicle-to-grid scheduling: Theory and algorithms}.
\newblock \bibinfo{journal}{\emph{IEEE Transactions on Intelligent
  Transportation Systems}} (\bibinfo{year}{2020}).
\newblock


\bibitem[\protect\citeauthoryear{Zhang, Zhou, Jiang, Wang, Zhang, and
  Chen}{Zhang et~al\mbox{.}}{2021}]%
        {zhang2020plug}
\bibfield{author}{\bibinfo{person}{Yanru Zhang}, \bibinfo{person}{Yingjie
  Zhou}, \bibinfo{person}{Changkun Jiang}, \bibinfo{person}{Yan Wang},
  \bibinfo{person}{Ruichang Zhang}, {and} \bibinfo{person}{George Chen}.}
  \bibinfo{year}{2021}\natexlab{}.
\newblock \showarticletitle{Plug-in Electric Vehicle Charging With Multiple
  Charging Options: A Systematic Analysis of Service Providers’ Pricing
  Strategies}.
\newblock \bibinfo{journal}{\emph{IEEE Transactions on Smart Grid}}
  \bibinfo{volume}{12}, \bibinfo{number}{1} (\bibinfo{year}{2021}),
  \bibinfo{pages}{524--537}.
\newblock


\bibitem[\protect\citeauthoryear{Zhang, Zhao, Tang, and Low}{Zhang
  et~al\mbox{.}}{2018}]%
        {zhang2016profit}
\bibfield{author}{\bibinfo{person}{Ying Jun~Angela Zhang},
  \bibinfo{person}{Changhong Zhao}, \bibinfo{person}{Wanrong Tang}, {and}
  \bibinfo{person}{Steven~H Low}.} \bibinfo{year}{2018}\natexlab{}.
\newblock \showarticletitle{Profit-maximizing planning and control of battery
  energy storage systems for primary frequency control}.
\newblock \bibinfo{journal}{\emph{IEEE Transactions on Smart Grid}}
  \bibinfo{volume}{9}, \bibinfo{number}{2} (\bibinfo{year}{2018}),
  \bibinfo{pages}{712--723}.
\newblock


\bibitem[\protect\citeauthoryear{Zhu and Zhang}{Zhu and Zhang}{2019}]%
        {zhu2018optimal}
\bibfield{author}{\bibinfo{person}{Diwei Zhu} {and}
  \bibinfo{person}{Ying-Jun~Angela Zhang}.} \bibinfo{year}{2019}\natexlab{}.
\newblock \showarticletitle{Optimal coordinated control of multiple battery
  energy storage systems for primary frequency regulation}.
\newblock \bibinfo{journal}{\emph{IEEE Transactions on Power Systems}}
  \bibinfo{volume}{34}, \bibinfo{number}{1} (\bibinfo{year}{2019}),
  \bibinfo{pages}{555--565}.
\newblock


\end{thebibliography}


\end{document}